\documentclass[journal]{IEEEtran}

\ifCLASSINFOpdf
\else
\fi
\usepackage{graphicx}
\usepackage{bm}
\usepackage{amsmath}
\usepackage{flushend}
\DeclareMathOperator{\sinc}{sinc}

\DeclareMathOperator{\Si}{Si}
\DeclareMathOperator{\Ci}{Ci}

\begin{document}

\title{Radiation from transmission lines PART II: insulated transmission lines}

\author{\IEEEauthorblockN{Reuven Ianconescu}
\IEEEauthorblockA{Department of Electrical Engineering\\
Shenkar College of Engineering and Design\\
12, Anna Frank St., Ramat Gan, Israel\\
Email: riancon@gmail.com}\\
\and
\IEEEauthorblockN{Vladimir Vulfin}
\IEEEauthorblockA{Department of Electrical and Computer Engineering\\
Ben-Gurion University of the Negev\\
Beer Sheva 84105, Israel\\
Email: vlad2042@yahoo.com}
}

\maketitle

\begin{abstract}
We develop in this work a radiation losses model for Quasi-TEM two-conductors transmission lines
insulated in a dielectric material. The analysis is exact, based on Maxwell equations and all the analytic
results are validated by comparison with ANSYS-HFSS simulation results and previous published works.
\end{abstract}


\section{Introduction}

We presented in \cite{full_model_arxiv} (and previous conferences \cite{eumw_2016,icsee_2016})
an analysis of radiation losses from two-conductors
transmission lines (TL) in free space, in which we analysed semi-infinite as well as finite
TL, and showed that the radiation from TL is essentially a termination phenomenon. We found
that the power radiated by a finite TL, carrying a forward current $I^+$, tends to the constant
$60 \Omega\, (kd)^2|I^+|^2$ ($k$ being the wavenumber and $d$ the effective separation between the
conductors) when the TL length tends to infinity (in practice overpasses several wavelengths).
This constant is twice the power radiated by a semi-infinite TL, showing that a very long TL
can be regarded as two separate semi-infinite TL, see \cite{full_model_arxiv}.

The purpose of this work is to generalise the results in \cite{full_model_arxiv} to Quasi-TEM
two-conductors TL isolated by a lossless dielectric material.
We remark that power loss from TL is also affected by nearby objects
interfering with the fields, line bends, irregularities, etc. This is certainly true, but those affect {\it not
only} the radiation, {\it but also} the basic, ``ideal'' TL model in what concerns the characteristic
impedance, the propagation wave number, etc. Like in \cite{full_model_arxiv} (and references therein) those non-ideal phenomena are
{\it not considered} in the current work, which derives the radiation-losses for ideal, non bending, fixed cross section TL.

\begin{figure}[!tbh]
\includegraphics[width=8cm]{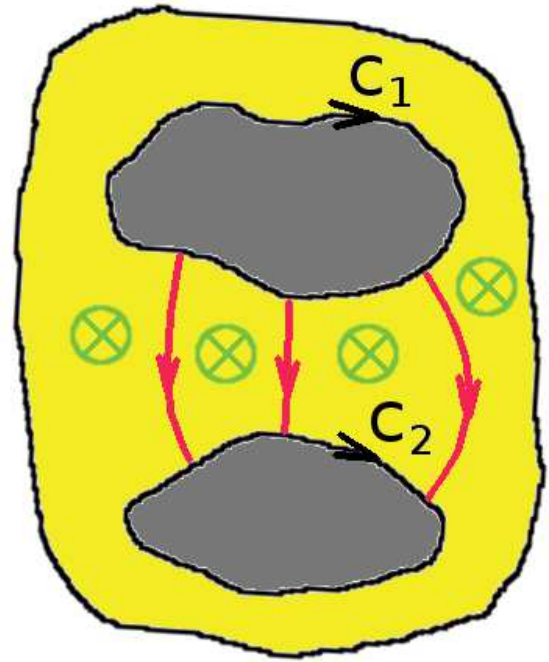}
\caption{A general cross section of two conductors insulated in a dielectric. The grey regions are
the ideal conductors and $c_{1,2}$ are the contours of those conductors. The dielectric
(yellow), is of uniform relative dielectric permittivity $\epsilon_r$. Under excitation the dielectric
insulator develops polarisation currents. The transverse polarisation current density (red arrows)
is $j\omega\epsilon_0(\epsilon_r-1)\mathbf{E}_T$, $\omega$ being the angular frequency and $\mathbf{E}_T$
the transverse component of the electric field. The longitudinal ($z$ directed) polarisation current density
(green) is $j\omega\epsilon_0(\epsilon_r-1)E_z$, $E_z$ being the $z$ component of the electric field.}
\label{config}
\end{figure}

The case of TL in dielectric insulator is much more complicated than the free space case. The fact that the TL propagation wavenumber $\beta$ is
different form the free space wavenumber $k$ by itself complicates the mathematics (see \cite{comcas_2017}),
but in addition it comes out that one needs to consider in this case polarisation currents in addition to free currents.
Hence, to define a generic algorithm for determining the radiation losses for two-conductors TL isolated in dielectric
material, one needs a generic specification for the polarisation currents. Given polarisation current elements
are summed vectorially (see red lines in Figure~\ref{config}), so that elements perpendicular to the vector sum do not contribute, one has to
define an average relative dielectric permittivity, named $\epsilon_{p}$ (the subscript ``p'' stands for polarisation), which is smaller or equal to the
known equivalent relative dielectric permittivity $\epsilon_{eq}$ \cite{orfanidis, pozar}.

\begin{figure}[!tbh]
\includegraphics[width=8cm]{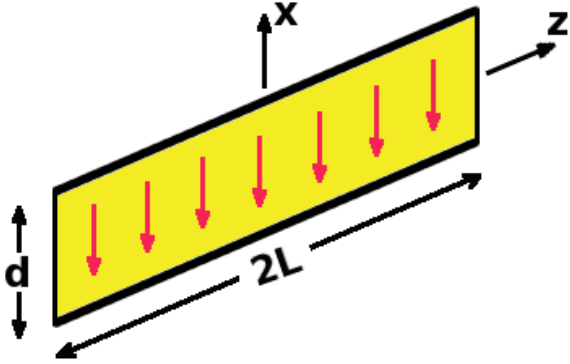}
\caption{Twin lead equivalent of TL insulated in dielectric material. The free TL currents in the
conductors at $x=\pm d/2$ are $\pm I^+ e^{-j\beta z}$ respectively, where the positive direction
is $\widehat{z}$. The free termination currents in the conductors at $z=\mp L$ are 
$\pm I^+e^{\pm j\beta L}$ respectively, where the positive direction is $\widehat{x}$. The polarisation currents
are represented as surface currents on the plane $y=0$. Their value is
$\mathbf{J}_p=-\widehat{x}\frac{\epsilon_{p}-1}{\epsilon_{p}}j\beta I^+\delta(y)$,
see Eqs.~(\ref{J_s_d1})-(\ref{J_p}), and the physical meaning of $\epsilon_{p}$ is explained in Appendix~C.}
\label{twin_lead}
\end{figure}

There are three appendices in this work. Appendix~A explains some basics on Quasi-TEM cross section
behaviour. We discuss the propagation wavenumber $\beta$, the equivalent relative dielectric permittivity
$\epsilon_{eq}$ and their connection to the capacitance per length unit $C$ and the characteristic
impedance $Z_0$ (which are strictly speaking well defined only for ``pure'' TEM). In Appendix~B we
develop the far potential vector, and similarly to \cite{full_model_arxiv}, we show that one can
represent any two-conductors TL isolated in dielectric material, by a twin lead (see Figure~\ref{twin_lead}), provided the
electrical size of the cross section is small.
As mentioned before, we calculate in the appendix
also the contribution of the polarisation currents, and those require a generic definition
of an average relative dielectric permittivity, $\epsilon_{p}$. The connection between $\epsilon_{p}$
and $\epsilon_{eq}$ and some more insight into their physical meaning is discussed in Appendix~C.

The main text is organised as follows.
In section~II we calculate the power radiated by a TL carrying a forward wave $I(z)=I^+e^{-j\beta z}$
(matched TL), for a general cross section of the TL, as shown in Figure~\ref{config}.
 We base the calculations on
the results of Appendix~B, in which we show (similarly to \cite{full_model_arxiv}) that the radiation
from a TL of any cross section of small electrical dimensions can be formulated in terms of a twin lead
analogue as shown in Figure~\ref{twin_lead}, but unlike in the free space case, this twin lead includes
also a sheet of polarisation surface current.
After deriving the radiated power and the radiation pattern for the matched TL, we show the limit of the free
space case and the limit of a long TL, and how this connects to a semi-infinite TL.
The results for the matched TL are generalised for any combination of waves $I(z)=I^+e^{-j\beta z}+I^-e^{j\beta z}$.

In section~III, we validate the theoretical results obtained in section~II, by comparing them
with a previous work dealing with radiation from TL \cite{Nakamura_2006}. This work was concerned with
reducing radiation losses by using a side plate (mirror) to create opposite image currents, and
they considered the free space case, and also TL inside dielectric, but ignored polarisation currents.
Reducing our configuration to the assumptions in \cite{Nakamura_2006}, shows a very good comparison
with these results. We then compare our theoretical results with ANSYS-HFSS commercial
software simulation results for two cross section examples. The work is ended by some concluding remarks.

Note: through this work, we use RMS values, hence there is no 1/2 in the expressions for power. Partial derivatives
are abbreviated, like for example derivative with respect to time $\frac{\partial}{\partial t}\equiv\partial_t$. 

 

\section{Radiated power}

\subsection{Matched TL}

We calculate in this section the power radiated from a matched general Quasi-TEM two-conductors TL insulated
in a dielectric material of any cross section, as shown in Figure~\ref{config}, carrying a forward wave described
by the current
\begin{equation}
I(z)=I^+e^{-j\beta z},
\label{I_z_matched}
\end{equation}
where $-L\le z \le L$. As shown in Appendix~B
(similarly to \cite{full_model_arxiv}), for the purpose of calculating far fields, any general cross
section, can be explored by its equivalent twin lead representation shown in Figure~\ref{twin_lead}. 
The free currents in the TL line conductors and longitudinal ($z$ directed) polarisation currents
define the separation distance $d$ between the conductors in the twin lead representation (Eq.~\ref{d}),
and define the $z$ component of the far potential vector computed in Eq.~(\ref{A_z}).

The free termination currents of the TL and the transverse polarisation current density, represented
by a sheet of $x$ directed surface polarisation current density $J_{s\, p}$ in the twin lead representation
define the $x$ directed component of the far potential vector computed in Eq.~(\ref{A_x}).

Using these results from Appendix~B, we calculate here the total radiated power from
the TL. Starting with the contribution of the longitudinal currents, we rewrite Eq.~(\ref{A_z})
in this form
\begin{equation}
A_z=\mu_0 G(r) F_{(z)}(\theta,\varphi)
\label{A_z_na}
\end{equation}
where
\begin{equation}
F_{(z)}(\theta,\varphi)\equiv jk 2L d I^+\sinc[kL(\cos\theta-n_{eq})] \sin\theta\cos\varphi,
\label{F_z}
\end{equation}
and the subscript $(z)$ denotes the contribution from the $z$ directed currents, and $F_{(z)}$ is the directivity
associated with this contribution. To obtain the far fields (those decaying like $1/r$), the $\boldsymbol{\nabla}$ operator
is approximated by $-jk\mathbf{\widehat{r}}$ and one obtains:
\begin{equation}
\mathbf{H}_{(z)}=\frac{1}{\mu_0}\boldsymbol{\nabla}\times(A_z\mathbf{\widehat{z}})=jkG(r)F_{(z)}(\theta,\varphi)\sin\theta\boldsymbol{\widehat{\varphi}}
\label{H_z_contrib}
\end{equation}
and the electric field associated with it is $\mathbf{E}_{(z)}=\eta_0\mathbf{H}_{(z)}\times\mathbf{\widehat{r}}$.

To calculate the contribution of the transverse ($x$ directed currents), we rewrite Eq.~(\ref{A_x})
in this form
\begin{equation}
A_x=\mu_0G(r) F_{(x)}(\theta,\varphi)
\label{A_x_na}
\end{equation}
where
\begin{equation}
F_{(x)} \equiv -jk I^+d 2L\sinc[kL(\cos\theta-n_{eq})](\cos\theta-n_{eq}/\epsilon_{p})
\label{F_x}
\end{equation}
and the subscript $(x)$ denotes the contribution from the $x$ directed currents, and $F_{(x)}$ is the directivity
associated with this contribution.

The parameter $\epsilon_{p}$ comes from defining the polarisation currents as $(\epsilon_{eq}-1)/\epsilon_{eq}$
times the displacement current, and as explained in Appendix~C, the ratio
$\frac{(\epsilon_{p}-1)/\epsilon_{p}}{(\epsilon_{eq}-1)/\epsilon_{eq}}$ represents the average projection
factor of the polarisation current elements on the main ($x$) axis - the axis with respect to which the twin
lead model has been defined (see Figure~\ref{twin_lead}). In cross sections having a transverse E field mainly
in the $x$ direction the projection factor is close to 1, hence $\epsilon_{p}\simeq\epsilon_{eq}$, and in the
opposite extreme case $\epsilon_{p}\simeq 1$ (negligible polarisation currents), so that
$1\le \epsilon_{p}\le\epsilon_{eq}=n_{eq}^2$. As evident from Eq.~(\ref{F_x}), $\epsilon_{p}$ always appears in the ratio
$n_{eq}/\epsilon_{p}$, we therefore use the definition
\begin{equation}
\overline{n}\equiv n_{eq}/\epsilon_{p},
\label{overline_n}
\end{equation}
so that
\begin{equation}
1/n_{eq}\le\overline{n}\le n_{eq}.
\label{overline_n_range}
\end{equation}
We may therefore use $\overline{n}=n_{eq}^a$, so that the power $a$ satisfies $-1\le a\le 1$, but as explained in Appendix~C, the equality
case $a=1$ is not physical, so it is considered {\it only} in the context of ``ignoring the transverse polarisation''.

To obtain the far fields, we use
$\mathbf{H}_{(x)}=\frac{1}{\mu_0}\boldsymbol{\nabla}\times(A_x\mathbf{\widehat{x}})$ and the
identity $\mathbf{\widehat{r}}\times\mathbf{\widehat{x}}= \cos\theta\cos\varphi\boldsymbol{\widehat{\varphi}} + 
\sin\varphi\boldsymbol{\widehat{\theta}}$, getting
\begin{equation}
\mathbf{H}_{(x)}=-jk[\cos\theta\cos\varphi\boldsymbol{\widehat{\varphi}} + \sin\varphi\boldsymbol{\widehat{\theta}}]G(r)F_{(x)}
\label{H_x_contrib}
\end{equation}
and the electric field associated with it is $\mathbf{E}_{(x)}=\eta_0\mathbf{H}_{(x)}\times\mathbf{\widehat{r}}$. 

\begin{figure}[!tbh]
\includegraphics[width=8cm]{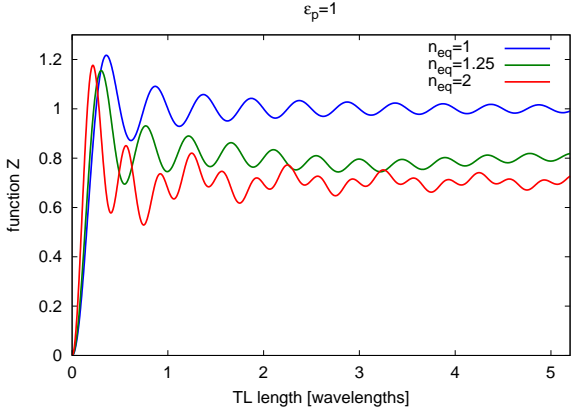}
\caption{$Z$ from Eq.~(\ref{Z}) as function of the TL line length in units of wavelengths, for
$\overline{n}=n_{eq}$, i.e. $\epsilon_{p}=1$ and for values of $n_{eq}=1$, 1.25 and 2. The asymptotic
values for a long TL are 1, 0.79 and 0.704 respectively (see Eq.~\ref{Z_long_TL}), and those asymptotic values tends to 2/3,
for big $n_{eq}$, according to the case $\overline{n}=n_{eq}$ in Eq.~(\ref{Z_long_TL_big_n}).}
\label{epsilon_av_1}
\end{figure}

\begin{figure}[!tbh]
\includegraphics[width=8cm]{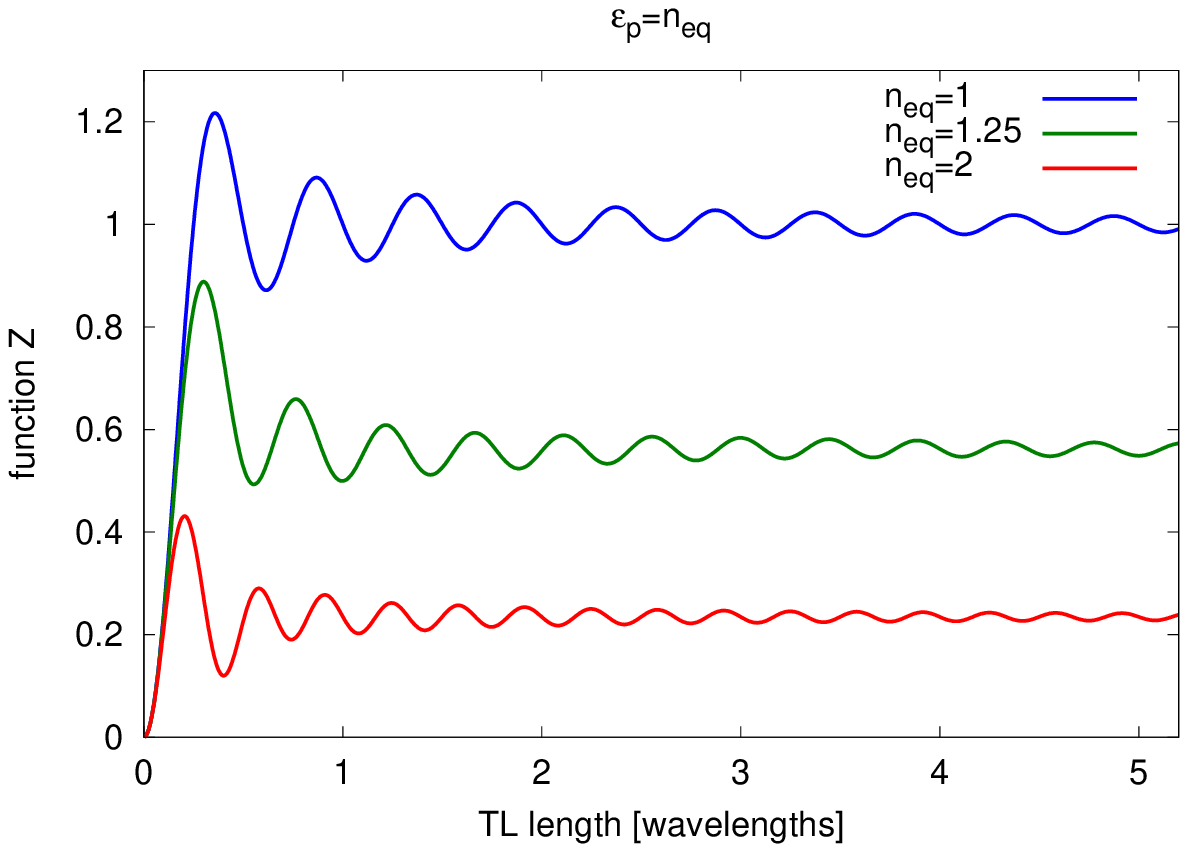}
\caption{Same as Figure~\ref{epsilon_av_1}, only for $\overline{n}=1$, i.e. $\epsilon_{p}=n_{eq}$.
The asymptotic values for a long TL are 1, 0.56 and 0.23 for $n_{eq}=1$, 1.25 and 2, respectively
(see Eq.~\ref{Z_long_TL}), and those asymptotic values tends to 0, for big $n_{eq}$, according
to the case $\overline{n}\neq n_{eq}$ in Eq.~(\ref{Z_long_TL_big_n}).}
\label{epsilon_av_n_eq}
\end{figure}

\begin{figure}[!tbh]
\includegraphics[width=8cm]{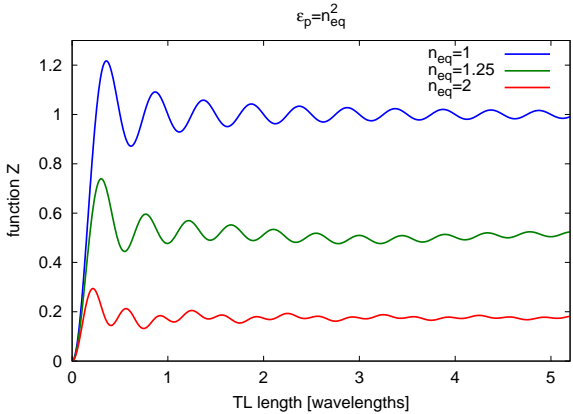}
\caption{Same as Figure~\ref{epsilon_av_1}, only for $\overline{n}=1/n_{eq}$, i.e. $\epsilon_{p}=\epsilon_{eq}=n_{eq}^2$.
The asymptotic values for a long TL are 1, 0.5 and 0.18 for $n_{eq}=1$, 1.25 and 2, respectively
(see Eq.~\ref{Z_long_TL}), and those asymptotic values tends to 0, for big $n_{eq}$, according to the
case $\overline{n}\neq n_{eq}$ in Eq.~(\ref{Z_long_TL_big_n}).}
\label{epsilon_av_n_eq_squared}
\end{figure}

Now summing Eqs.~(\ref{H_z_contrib}) with (\ref{H_x_contrib}) we obtain the total far magnetic field
\begin{align}
  \mathbf{H}^+=&-k^2G(r)I^+d 2L\sinc[kL(\cos\theta-n_{eq})] \notag \\
             &[\boldsymbol{\widehat{\theta}}\sin\varphi(\cos\theta-\overline{n})+\boldsymbol{\widehat{\varphi}}\cos\varphi(1-\overline{n}\cos\theta)]
\label{H}
\end{align}
and the electric field $\mathbf{E}^+=\eta_0\mathbf{H}^+\times\mathbf{\widehat{r}}$. We use from here the superscript ``+'',
because those results are for a forward wave. The Poynting vector is
\begin{align}
S^+=\eta_0|\mathbf{H}^+|^2=&\frac{\eta_0 k^4 |I^+|^2 d^2 L^2}{4\pi^2r^2}\sinc^2[kL(\cos\theta-n_{eq})] \notag \\
             &[\sin^2\varphi(\cos\theta-\overline{n})^2+\cos^2\varphi(1-\overline{n}\cos\theta)^2],
\label{S}
\end{align}
and the total radiated power is calculated via
\begin{equation}
P^+_{rad}=\int_0^{2\pi}\int_0^{\pi}\sin\theta d\theta d\varphi r^2 S^+.
\label{P_rad_1}
\end{equation}
We remark that $\int_0^{2\pi}d\varphi\sin^2\varphi=\int_0^{2\pi}d\varphi\cos^2\varphi=\pi$, so that
the radiated power is given by the single integral in $\theta$. After changing variable: $y=-\cos\theta$,
one obtains
\begin{align}
P^+_{rad}=&60\,\Omega (kd)^2|I^+|^2\int_{-1}^{1}dy\,(kL)^2 \sinc^2[kL(n_{eq}+y)] \notag \\
  &[(1+\overline{n}^2)(1+y^2)/2+2 \overline{n}y]
\label{P_rad_2}
\end{align}
The integral is carried out analytically, resulting in an expression which is very big, and therefore
we introduce some definitions. We define the following function arguments:
\begin{equation}
a_+\equiv 2kL(n_{eq}+1) \,\,;\,\, a_-\equiv 2kL(n_{eq}-1),
\label{a_pm}
\end{equation}
Furthermore, we define
\begin{equation}
Q\equiv \frac{\cos(a_+)}{a_+} - \frac{\cos(a_-)}{a_-} + \Si(a_+) - \Si(a_-)
\label{QQQ}
\end{equation}
\begin{equation}
W\equiv \ln\frac{n_{eq}+1}{n_{eq}-1}   - [\Ci(a_+)-\Ci(a_-)]
\label{WWW}
\end{equation}
where $\Si$ and $\Ci$ are the sine and cosine integral functions respectively. The solution of the
$(1+y^2)$ part in the integral in Eq.~(\ref{P_rad_2}), without the prefactor $(1+\overline{n}^2)/2$, is given
by the function $Z_1$:
\begin{align}
Z_1(kL,n_{eq})=&\frac{2n^2_{eq}}{n^2_{eq}-1}+ kL(n^2_{eq}+1)Q  - n_{eq}W - \notag \\
               & \frac{\sin(a_+)-\sin(a_-)}{4kL}
\label{Z1}
\end{align}
and the solution of the $y$ part in the integral in Eq.~(\ref{P_rad_2}), without the prefactor $2\overline{n}$, is given
by the function $Z_2$ as follows:
\begin{equation}
Z_2(kL,n_{eq})=\frac{-n_{eq}}{n^2_{eq}-1} - kLn_{eq}Q + W/2 .
\label{Z2}
\end{equation}
So the solution of the whole integral is described by the function $Z(kL,n_{eq},\overline{n})$
\begin{equation}
Z(kL,n_{eq},\overline{n})=\frac{1+\overline{n}^2}{2} Z_1 + 2\overline{n} Z_2.
\label{Z}
\end{equation}
The behaviour of $Z(kL,n_{eq},\overline{n})$ is shown in Figures~\ref{epsilon_av_1}-\ref{epsilon_av_n_eq_squared}
and referred to hereinafter. Looking at the figures, we understand that as $n_{eq}$ is bigger, the function
$Z(kL,n_{eq},\overline{n})$ decreases, while for a given $n_{eq}$, bigger transverse polarisation currents
(bigger $\epsilon_{p}$, hence smaller $\overline{n}$), further decrease $Z(kL,n_{eq},\overline{n})$.

From (\ref{P_rad_2}) and (\ref{Z}), the expression for the radiated power is
\begin{equation}
P^+_{rad}= 60\,\Omega |I^+|^2(kd)^2 Z(kL,n_{eq},\overline{n})
\label{P_rad_3}
\end{equation}
The radiation pattern function is calculated from the radial pointing vector (Eq.~(\ref{S})) and the total power
in Eq.~(\ref{P_rad_3}), using $D^+=4\pi r^2 S^+/P^+_{rad}$ which comes out
\begin{align}
D^+&(\theta,\varphi)=2 \sin^2[kL(\cos\theta-n_{eq})] \notag \\
&\frac{\sin^2\varphi(\cos\theta-\overline{n})^2+\cos^2\varphi(1-\overline{n}\cos\theta)^2}{Z(kL,n_{eq},\overline{n}) (\cos\theta-n_{eq})^2}
\label{D}
\end{align}
and the radiated power relative to the forward wave propagating power ($P^+=|I^+|^2Z_0$) is given by
\begin{equation}
\frac{P^+_{rad}}{P^+}=\frac{60\,\Omega}{Z_0}(kd)^2 Z(kL,n_{eq},\overline{n}),
\label{rel_P_rad}
\end{equation}

The expressions for the radiated power and radiation pattern are complicated (certainly relative to the
free space case \cite{full_model_arxiv}) and it would be of interest
to compare them to the free space case and determine some limits, in the following subsections.

\subsubsection{The free space limit}

In this limit $n_{eq}=1$, and also $\overline{n}=1$, according to Eq.~(\ref{overline_n_range}).
So $a_+=4kL$, $a_-=0$ and $Z$ in Eq.~(\ref{Z}) is $Z_1+2Z_2$, resulting in
\begin{equation}
Z= \frac{2n^2_{eq}}{n^2_{eq}-1}+ 2kLQ - W - \frac{\sin(4kL)}{4kL}+ 2\frac{-n_{eq}}{n^2_{eq}-1} - 2kLQ + W,
\label{Z_free_space}
\end{equation}
We note that
\begin{equation}
2\frac{n^2_{eq}-n_{eq}}{n^2_{eq}-1}=2\frac{n_{eq}(n_{eq}-1)}{(n_{eq}-1)(n_{eq}+1)}\rightarrow 1
\label{limit}
\end{equation}
for $n_{eq}\rightarrow 1$, so we recover the free space formula for $Z$
\begin{equation}
Z(kL,n_{eq}=1,\overline{n}=1)= 1 - \sinc(4kL),
\label{Z_free_space1}
\end{equation}
shown in blue colour in Figures~\ref{epsilon_av_1}-\ref{epsilon_av_n_eq_squared}, see \cite{full_model_arxiv}.
The limit of the radiation pattern is
\begin{align}
D^+(\theta,\varphi)=2& \sin^2[kL(1-\cos\theta)] \notag \\
&\frac{(1-\cos\theta)^2(\sin^2\varphi+\cos^2\varphi)}{(1 - \sinc(4kL)) (1-\cos\theta)^2},
\label{D_free_space}
\end{align}
which remains only a function of $\theta$, recovering Eq.~(12) in \cite{full_model_arxiv}.

\subsubsection{The limit of a long TL}

In the free space case the limit for a long TL ($kL\rightarrow\infty$) is simply $Z=1$. In our
case this limit depends on $n_{eq}$ and $\overline{n}$. For $kL\rightarrow\infty$, $a_+$ and
$a_-$ both go to $\infty$. The $\Ci$ function goes to 0 for large argument, therefore Eq.~(\ref{WWW})
reduces to
\begin{equation}
W \rightarrow \ln\frac{n_{eq}+1}{n_{eq}-1},
\label{WWW_long_TL}
\end{equation}
The function $Q\rightarrow 0$, but here one has to be careful, because we need the limit of
$kLQ$ in Eqs.~(\ref{Z1}) and (\ref{Z2}). The $\Si$ function for large argument behaves like: 
\begin{align}
\Si(x)=\frac{\pi}{2}&-\frac{\cos x}{x}\left(1-\frac{2!}{x^2}+\frac{4!}{x^4}-\frac{6!}{x^6}\cdots\right)- \notag \\
                    &\frac{\sin x}{x}\left(\frac{1}{x}-\frac{3!}{x^3}+\frac{5!}{x^5}-\frac{7!}{x^7}\cdots\right)
\label{Si_big_arg}
\end{align}
and from here it is easy to find that $Q$ decreases faster than $(kL)^{-1}$, hence
\begin{equation}
kLQ \rightarrow 0
\label{QQQ_long_TL}
\end{equation}
also $[\sin(a_+)-\sin(a_-)]/(4kL) \rightarrow 0$ for $kL\rightarrow\infty$, we therefore obtain
after some algebra the following limit for $Z$:
\begin{align}
Z(kL\rightarrow\infty,n_{eq},\overline{n})=&\frac{n_{eq}}{n^2_{eq}-1}\left[(1+\overline{n}^2)n_{eq}-2\overline{n}\right]+ \notag \\
&\left[\overline{n}-\frac{n_{eq}(1+\overline{n}^2)}{2}\right]\ln\frac{n_{eq}+1}{n_{eq}-1}
\label{Z_long_TL}
\end{align}
Those limits can be calculated for the cases shown in Figures~\ref{epsilon_av_1}-\ref{epsilon_av_n_eq_squared}
and yield 1, 0.79 and 0.704 for $\overline{n}=n_{eq}$ (Figure~\ref{epsilon_av_1}), 1, 0.56 and 0.23
for $\overline{n}=1$ (Figure~\ref{epsilon_av_n_eq}) and 1, 0.5 and 0.18 for $\overline{n}=1/n_{eq}$
(Figure~\ref{epsilon_av_n_eq_squared}), for the values of $n_{eq}=1$, 1.25 and 2, respectively.

As shown in \cite{full_model_arxiv}, this limit represents the radiation of a long TL carrying a forward wave,
so that:
\begin{equation}
P^+_{\text{rad (long TL)}}= 60\,\Omega |I^+|^2(kd)^2 Z(kL\rightarrow\infty,n_{eq},\overline{n}).
\label{P_rad_long_TL}
\end{equation}
which also corresponds to twice the power radiated from a semi-infinite TL. This means that the
power radiated by a semi-infinite TL carrying a forward wave is
\begin{equation}
P^+_{\text{rad (semi-infinite)}}= 30\,\Omega |I^+|^2(kd)^2 Z(kL\rightarrow\infty,n_{eq},\overline{n}),
\label{P_rad_semi-infinte}
\end{equation}
see Figure~9 in \cite{full_model_arxiv}.

\subsubsection{The limit of big relative dielectric permittivity, for long TL}

This limit is discussed in the context of a long TL, so the limit of (\ref{Z_long_TL}) for $n_{eq}\rightarrow\infty$
depends on $\overline{n}$, as follows:
\begin{equation}
Z(kL\rightarrow\infty,n_{eq}\rightarrow\infty,\overline{n})=
\left \{
  \begin{tabular}{cc}
  2/3 & \,\,\,\,$\overline{n}=n_{eq}$  \\
  0 &  \,\,\,\,$1/n_{eq}\le \overline{n} < n_{eq}$
  \end{tabular}
\right . ,
\label{Z_long_TL_big_n}
\end{equation}
so that there is a singular case of ``ignoring the transverse polarisation'', for which the limit is 2/3, as shown
in Figure~\ref{epsilon_av_1}, and for any practical case the limit is 0, meaning that the radiated power vanishes
for strong relative permittivity of the dielectric insulator $n_{eq}\rightarrow\infty$
(see Figures~\ref{epsilon_av_n_eq} and \ref{epsilon_av_n_eq_squared}).

\subsection{Generalisation for non matched TL}

We generalise here the result (\ref{P_rad_3}) obtained for the losses of a finite TL carrying
a forward wave to any combination of waves, as follows:
\begin{equation}
I(z)=I^+e^{-jkz}+I^-e^{jkz}
\label{for_back_current}
\end{equation}
where $I^+$ is the forward wave phasor current, as used in the previous subsection and $I^-$
is the backward wave phasor current, still defined to the right in the ``upper'' line in
Figure~\ref{twin_lead}.

The solution for the general current is obtained as superposition of the solutions for the fields
generated by $I^+e^{-jkz}$ and $I^-e^{jkz}$.
The solution for the backward moving wave $I^-e^{jkz}$, can be found by first solving for a {\it reversed} $z$
axis in Figure~\ref{twin_lead}, i.e. a $z$ axis going to the left, and replacing in the solution (\ref{H}) $I^+\rightarrow -I^-$,
so one obtains
\begin{align}
  \mathbf{H}^-=&-k^2G(r)(-I^-)d 2L\sinc[kL(\cos\theta'-n_{eq})] \notag \\
             &[\boldsymbol{\widehat{\theta}}'\sin\varphi'(\cos\theta'-\overline{n})+\boldsymbol{\widehat{\varphi}}'\cos\varphi'(1-\cos\theta'\, \overline{n})]
\label{H_minus}
\end{align}
where $\theta'$ and $\varphi'$ are the spherical angles for the reversed $z$ axis.
Now to express the solution for the backward wave in the original coordinates, defined by the
right directed $z$ axis, one has to replace: $\theta'=\pi-\theta$, $\varphi'=-\varphi$,
and therefore also $\boldsymbol{\widehat{\theta}}'= -\boldsymbol{\widehat{\theta}}$ and
$\boldsymbol{\widehat{\varphi}}'= -\boldsymbol{\widehat{\varphi}}$, and sum (\ref{H}) with (\ref{H_minus}),
obtaining
\begin{align}
  \mathbf{H}=&\mathbf{H}^++\mathbf{H}^-=-k^2G(r)2dL \notag \\
             &\{\boldsymbol{\widehat{\theta}}\sin\varphi[I^+\sinc[kL(\cos\theta-n_{eq})](\cos\theta-\overline{n})+ \notag \\
             &I^-\sinc[kL(\cos\theta+n_{eq})](\cos\theta+\overline{n})]+ \notag \\
             &\boldsymbol{\widehat{\varphi}}\cos\varphi[I^+\sinc[kL(\cos\theta-n_{eq})](1-\overline{n}\cos\theta)+ \notag \\
             &I^-\sinc[kL(\cos\theta+n_{eq})](1+\overline{n}\cos\theta)]\},
\label{H_plus_and_minus}
\end{align}
from which the electric field is $\mathbf{E}=\eta_0\mathbf{H}\times\mathbf{\widehat{r}}$, so that the Poynting
vector is
\begin{align}
&S=\eta_0|\mathbf{H}|^2=\frac{\eta_0 k^4 d^2 L^2}{4\pi^2r^2}\{\notag \\
  &|I^+|^2\sinc^2[kL(\cos\theta-n_{eq})][(a_-)^2\sin^2\varphi+(B^-)^2\cos^2\varphi]+ \notag \\
  &|I^-|^2\sinc^2[kL(\cos\theta+n_{eq})][(a_+)^2\sin^2\varphi+(B^+)^2\cos^2\varphi]+ \notag \\
  &2\Re\{I^+I^{-*}\}\sinc[kL(\cos\theta-n_{eq})]\sinc[kL(\cos\theta+n_{eq})] \notag \\
  &(a_-a_+\sin^2\varphi + B^-B^+\cos^2\varphi) \}
\label{S_plus_and_minus}
\end{align}
where we used the abbreviations: $A^{\pm}=\cos\theta\pm \overline{n}$ and
$B^{\pm}=1\pm\overline{n}\cos\theta$.

We calculate the radiated power using Eq.~(\ref{P_rad_1}), and obtain
\begin{equation}
P_{rad}=P_{rad}^+ + P_{rad}^- + P_{rad\,\,\,mix}
\label{P_rad_plus_and_minus}
\end{equation}
where $P_{rad}^{\pm}$ are the powers radiated by the individual forward and backward waves, and are
given by (\ref{P_rad_3}), using the adequate current:
\begin{equation}
P_{rad}^{\pm}= 60\,\Omega (kd)^2 |I^{\pm}|^2 Z(kL,n_{eq},\overline{n}),
\label{P_rad_plus_or_minus}
\end{equation}
and $P_{rad\,\,\,mix}$ is the power radiated by the interference between $I^+$ and $I^-$, and is given by
\begin{align}
P_{rad\,\,\,mix}= 60\,\Omega (kd)^2 \Re\{I^+I^{-*}\} \left[1-\overline{n}^2\right] Z_{mix}
\label{P_rad_mix}
\end{align}
where
\begin{align}
Z_{mix}&=\cos(2kLn_{eq})[1-\frac{n_{eq}+1/n_{eq}}{2}W]+ \notag \\
& \frac{n_{eq}+1/n_{eq}}{2}\sin(2kLn_{eq})[\Si(a_+) - \Si(a_-)]-\sinc(2kL).
\label{Z_mix}
\end{align}
The arguments $a_+$ and $a_-$ are defined in Eq.~(\ref{a_pm}) and $W$ is defined in Eq.~(\ref{WWW_long_TL}).

Unlike the free space case \cite{full_model_arxiv} in which $P_{rad\,\,\,mix}=0$, for TL in insulated dielectric
the interference between the waves contributes to the radiation, and of course the contribution vanishes in
the free space limit for which $\overline{n}=1$. To be mentioned that $\overline{n}=1$, may also occur in
the insulated case ($n_{eq}>1$) if $\epsilon_{p}=n_{eq}$.

In the next section we validate the analytic results obtained in this section, using ANSYS commercial
software simulation and additional published results on radiation losses from TL.

\section{Validation of the analytic results}
\subsection{Comparison with \cite{Nakamura_2006}}

In 2006 Nakamura et. al. published the paper ``Radiation Characteristics of a Transmission Line
with a Side Plate'' \cite{Nakamura_2006} which intends to reduce radiation losses from TLs using
a side plate. The side plate is a perfect conductor put aside the transmission line, to create
opposite image currents, and hence reduce the radiation.

The authors first derived the radiation from a TL without the side plate, for the free space
case, and also for TL inside dielectric, but the dielectric has been taken into account in what
concerns the propagation constant $\beta=n_{eq}k$ only, {\it ignoring} the polarisation currents.

Therefore for the sake of comparison with \cite{Nakamura_2006} we have to use $\epsilon_{p}=1$,
hence $\overline{n}=n_{eq}$ in all our results.

We first remark that $I_0$ in \cite{Nakamura_2006} is a forward current, and from Eq.~(19) in
\cite{Nakamura_2006}, it is evident that they used RMS values. Hence $I_0$ in \cite{Nakamura_2006}
is the equivalent of our $|I^+|$. Also they used (capital) $K$ for the equivalent refraction
index, called in this work $n_{eq}$.

In \cite{Nakamura_2006} they did not obtain analytic expression for the radiation as function
of TL length, but they did obtain analytic expressions for the long TL limit, with which we
compare here our results. Eq.~(\ref{Z_long_TL}) simplifies for $\overline{n}=n_{eq}$ to:
\begin{equation}
Z(kL\rightarrow\infty,n_{eq},\overline{n}=n_{eq})=n^2_{eq}-n_{eq}\frac{n^2_{eq}-1}{2}\ln\frac{n_{eq}+1}{n_{eq}-1}
\label{Z_long_TL_e_av1}
\end{equation}
so that the power radiated by a semi-infinite TL in Eq~(\ref{P_rad_semi-infinte}) is:
\begin{equation}
P^+_{\text{rad (semi-infinite)}}= 30\,\Omega |I^+|^2(kd)^2\left\{n^2_{eq}-n_{eq}\frac{n^2_{eq}-1}{2}\ln\frac{n_{eq}+1}{n_{eq}-1}\right\}
\label{P_rad_semi-infinte_nakamura}
\end{equation}
which is {\it exactly} what they called ``the radiated power from the input end (or output
end) alone'', given in Eq.~(30) in \cite{Nakamura_2006} (note that they used the distance between
conductors $2h$, corresponding to our $d$, from there the factor 4).

Also note that the limit of $Z$ in Eq.~(\ref{Z_long_TL_e_av1}) for $n_{eq}\rightarrow\infty$ is 2/3, according
to the case $\overline{n}=n_{eq}$ in Eq.~(\ref{Z_long_TL_big_n}). This may be confirmed by comparing Eqs.~(31)
and (32) in \cite{Nakamura_2006}.

\begin{figure}[!tbh]
\includegraphics[width=4.3cm]{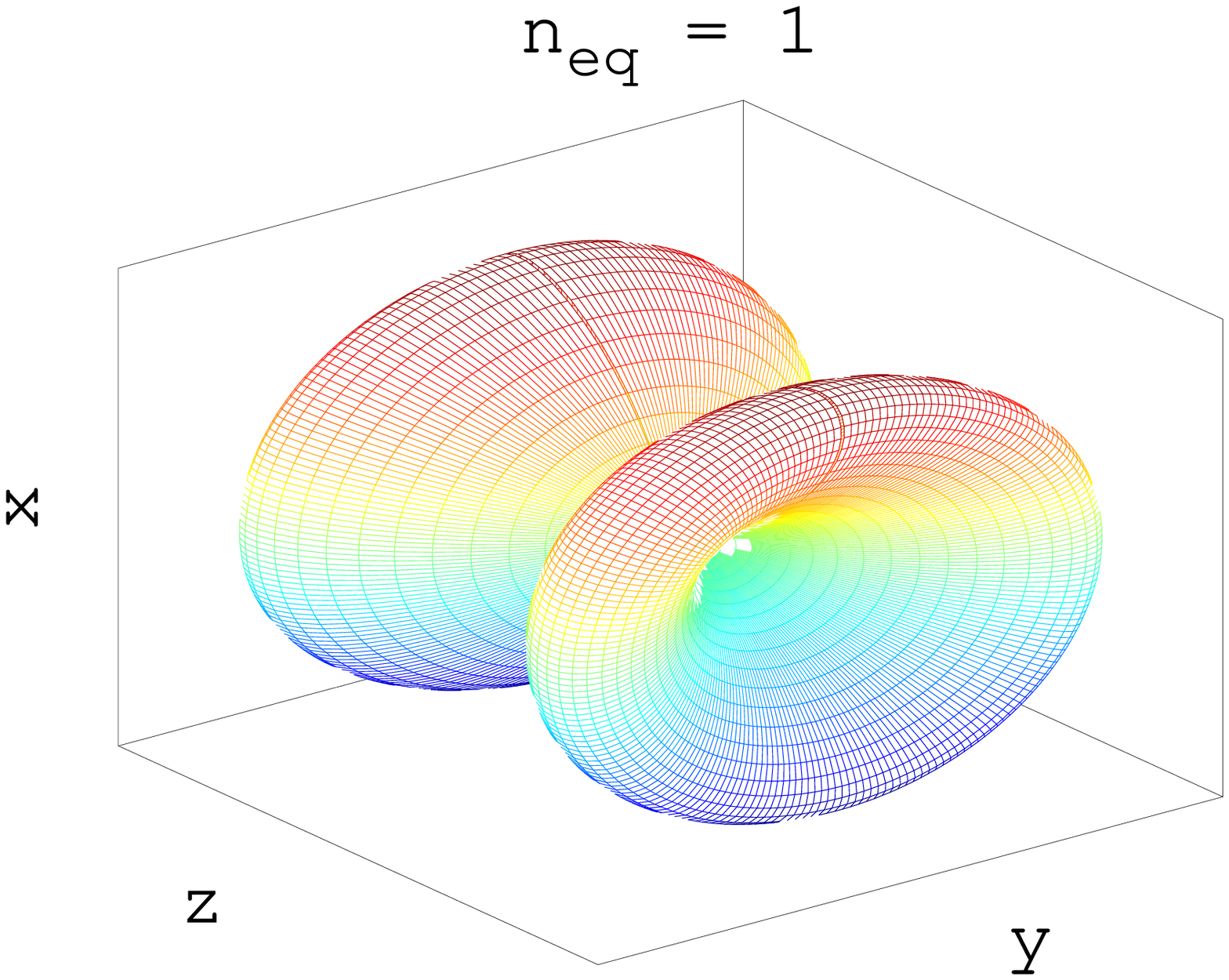}
\includegraphics[width=4.3cm]{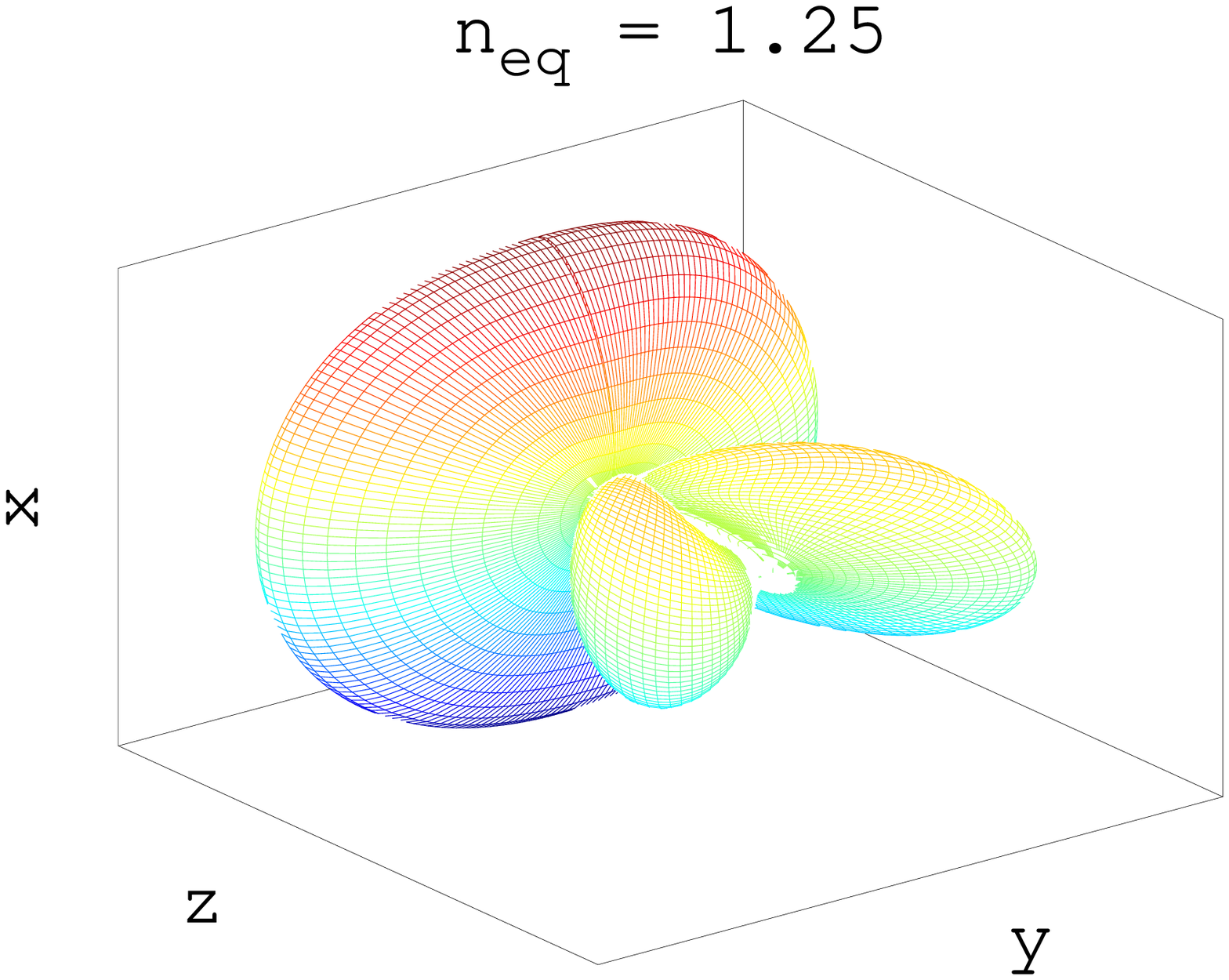}
\includegraphics[width=4.3cm]{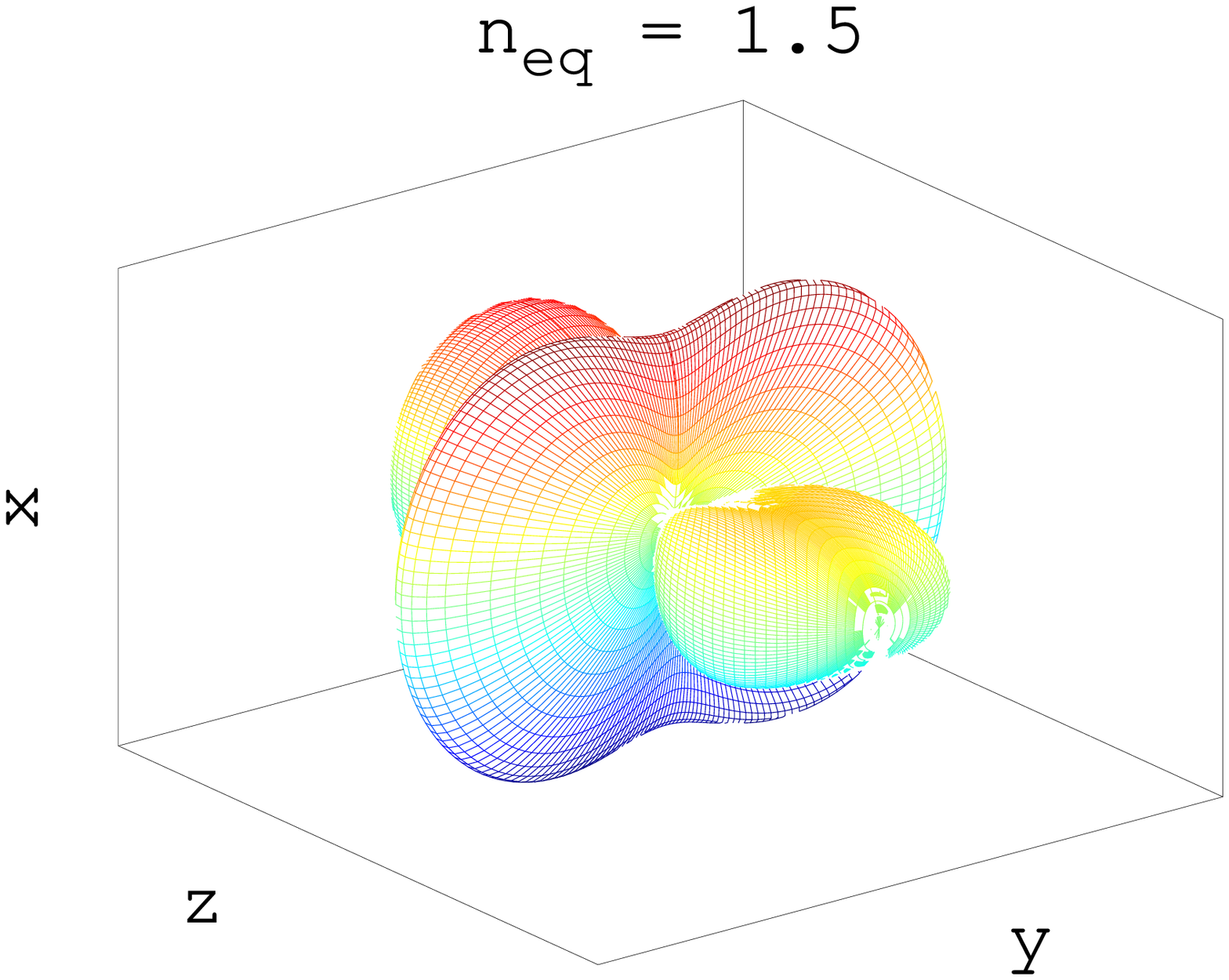}
\includegraphics[width=4.3cm]{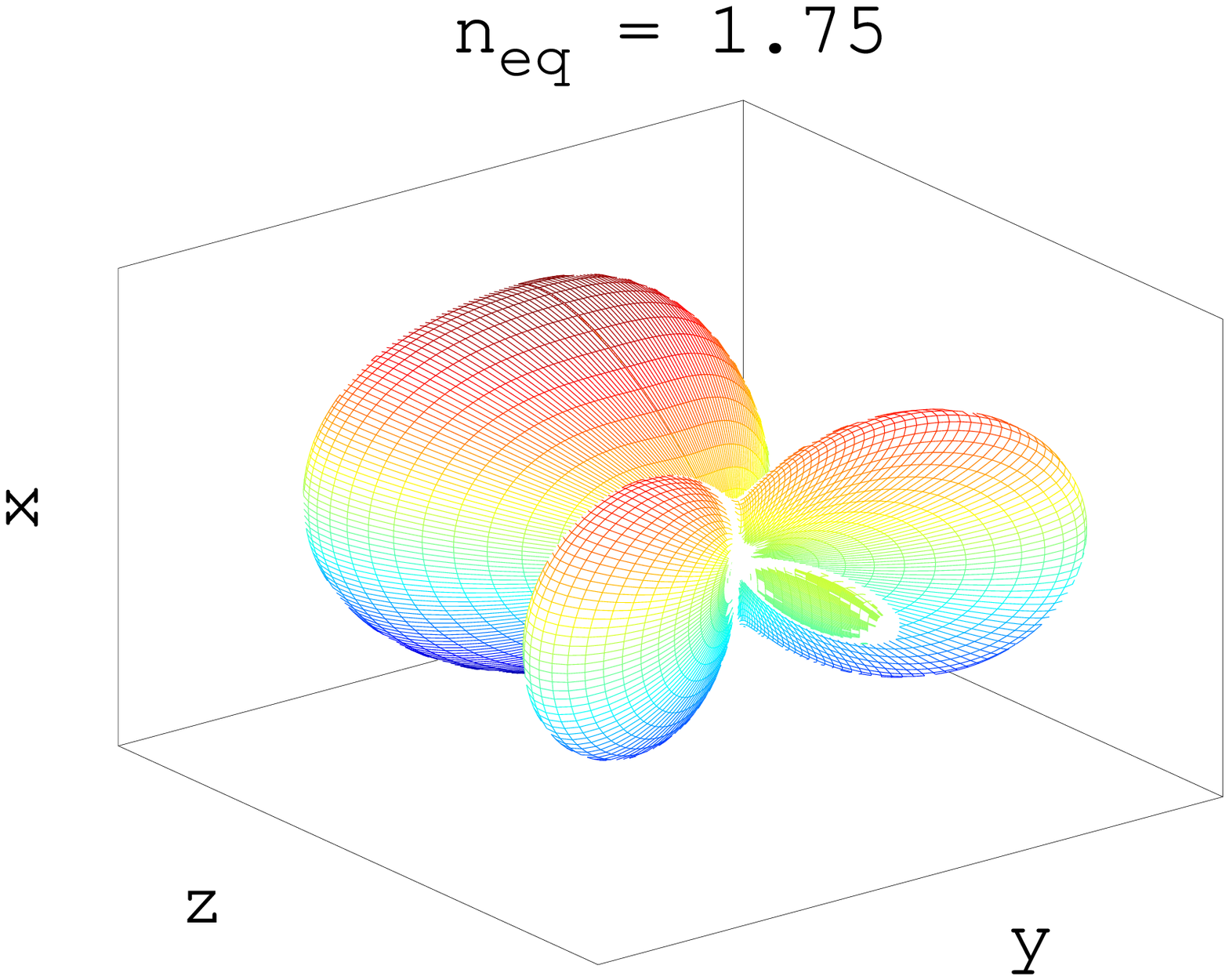}
\includegraphics[width=4.3cm]{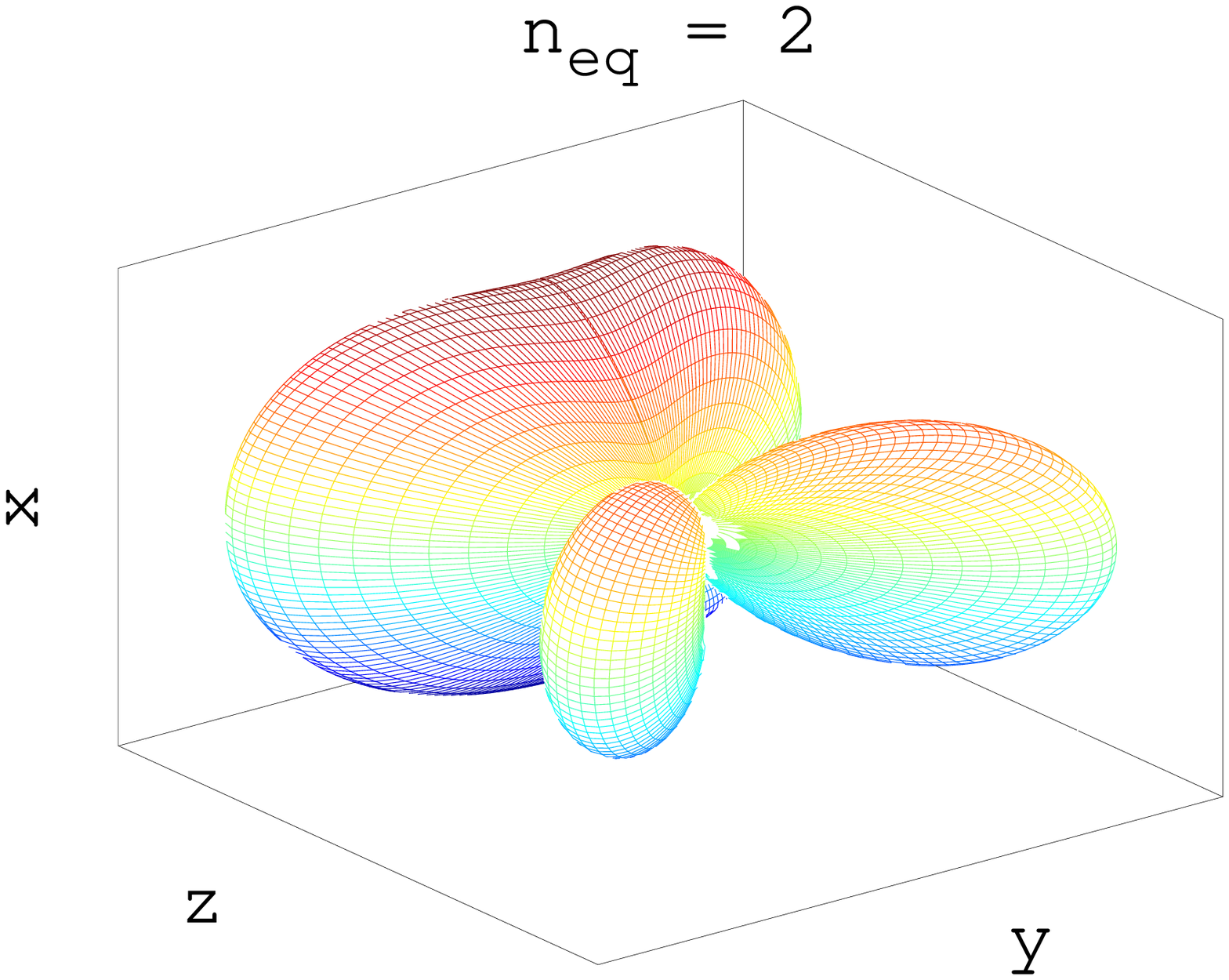}
\caption{Radiation pattern calculated from Eq.~(\ref{D}) with $\overline{n}=n_{eq}$ for TL of 1 wavelength for the cases $n_{eq}=1$, 1.25,
1.5, 1.75 and 2. They are identical to the parallel cases shown in Figure~5(b) of \cite{Nakamura_2006}. Note
that the definitions of the $x$ and $z$ axes are swapped in \cite{Nakamura_2006} compared to our definitions, we
therefore showed them in an orientation which makes the comparison easy (i.e. our $z$ axis is oriented in the
plots in the same direction as their $x$ axis).}
\label{rad_patt}
\end{figure}

Next we compare the radiation patterns obtained in Figure~5 of \cite{Nakamura_2006}, with ours. In
\cite{full_model_arxiv} we compared the free space case in panel (a), and here we compare our result
(Eq.~(\ref{D}) with $\overline{n}=n_{eq}$) with panel (b) of Figure~5 in \cite{Nakamura_2006}, showing the radiation patterns for a
TL of 1 wavelength, for different values of $n_{eq}$. This is shown in Figure~\ref{rad_patt}.
It is worthwhile to remark that the radiation pattern (Eq.~\ref{D}) does not depend on the distance between
the conductors $d$ (or $2h$ in \cite{Nakamura_2006}), hence the annotation of $h/\lambda=0.1$ in Figure~5 of 
\cite{Nakamura_2006} is redundant, and probably has been added to the caption because the authors computed
the radiation patterns numerically for $h/\lambda=0.1$, without deriving an analytic expression.

\begin{figure}[!tbh]
\includegraphics[width=9cm]{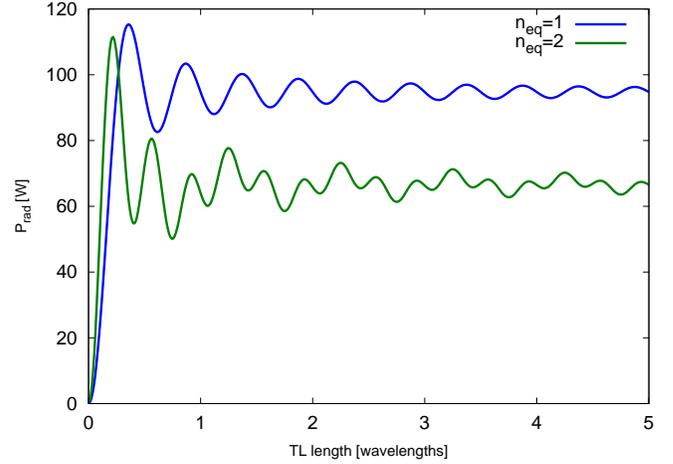}
\caption{Reproduction of Figure~6 in \cite{Nakamura_2006}, obtained by setting $60\,\Omega (kd)^2|I^+|^2=94.746$~[W]
and $\overline{n}=n_{eq}$ in Eq.~(\ref{P_rad_3}). One can check with an image editor that the blue and green
lines in this figure
{\it completely overlap} the solid and dashed lines in Figure~6 of \cite{Nakamura_2006}, respectively
(as shown also in \cite{comcas_2017}).}
\label{comparison_nakamura}
\end{figure}

Next, we compare our results with Figure~6 in \cite{Nakamura_2006}, which is the numerical integration
of Eq.~20 in \cite{Nakamura_2006} for the cases $n_{eq}=1$ and 2 (named $K=1$,2) where the solid line represents
the free space case ($n_{eq}=1$) and the dashed line represents the $n_{eq}=2$ case.
To calculate the result in Figure~6 of \cite{Nakamura_2006} they used $I_0=1$A, hence we set $|I^+|=1$A in
Eq.~(\ref{P_rad_3}). $2h$ is the distance between the conductors in ~\cite{Nakamura_2006},
equivalent to $d$ in this work, and they
used $h\lambda=0.1$, therefore $(kd)^2=(4\pi h/\lambda)^2=1.5791$ in Eq.~(\ref{P_rad_3}).
Hence the prefactor $60\,\Omega (kd)^2|I_0|^2=94.746$~[W]. The asymptotic value of $Z$ in Eq.~(\ref{Z_long_TL_e_av1})
is 1 for $n_{eq}=1$ and 0.704 for $n_{eq}=2$ (as shown also in Figure~\ref{epsilon_av_1}), therefore the
asymptotic power for $kL\gg 1$, for the two cases
shown in the figure are $94.75$~[W] and $94.75\times 0.704=66.72$~[W], respectively.
The results of Eq.~(\ref{P_rad_3}) for $n_{eq}=1$ and 2 are displayed in Figure~\ref{comparison_nakamura}
(and they are identical in shape to the corresponding cases of $n_{eq}=1$ and 2 in Figure~\ref{epsilon_av_1},
up to the constant $94.75$~[W]).

In the next sections we take two examples of cross section geometries on which we apply the analytic result
Eqs.~(\ref{P_rad_3}), (\ref{rel_P_rad}) or (\ref{P_rad_plus_and_minus}) and compare the results with simulation results of ANSYS-HFSS
commercial software, in the frequency domain, FEM technique. 

\subsection{Comparison with ANSYS simulation results - Example 1}

\begin{figure}[!tbh]
\includegraphics[width=8cm]{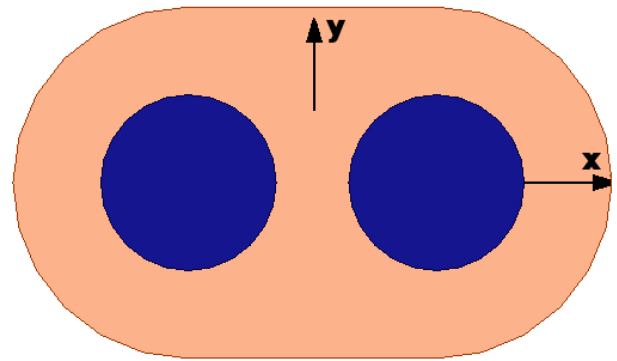}
\caption{The cross section consists of two circular shaped ideal conductors of radius $a=1.27$~cm (dark blue), the distance
between their centres being $s=3.59$~cm. The dielectric insulator (pink) is circular with radius $2a$ for $|x|>s/2$
and rectangular in the region $|x|<s/2$. The relative permittivity of the dielectric insulator is $\epsilon_r=3$.}
\label{cross_section_ins}
\end{figure}

\begin{figure}[!tbh]
\includegraphics[width=8cm]{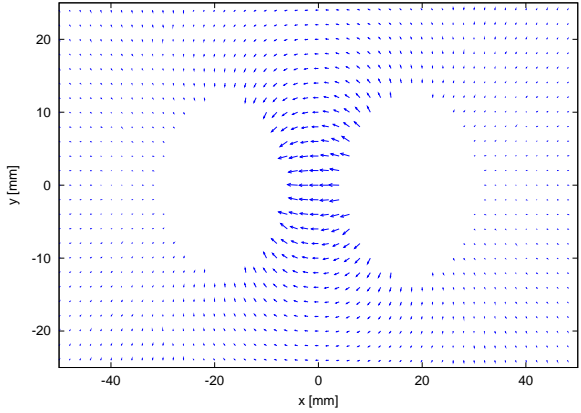}
\caption{Arrow plot of $\mathbf{E}_T$ for the cross section shown in Figure~\ref{cross_section_ins}.}
\label{quiver_cylinders}
\end{figure}

In this example we use the cross section shown in Figure~\ref{cross_section_ins}, which is
similar to the one used in \cite{full_model_arxiv}, only insulated in a dielectric
material.
We performed an ANSYS-HFSS cross section analysis at the frequency 240~MHz. From this analysis
we obtained the propagation constant $\beta=n_{eq}k=8.1$~[1/m], establishing the equivalent
refraction index $n_{eq}=1.613$.
An arrow plot of the transverse electric field $\mathbf{E}_T$ is shown in Figure~\ref{quiver_cylinders}.

From this analysis, using Eqs.~(\ref{d0_f}), (\ref{alpha_f} and (\ref{d1_f}) one finds that
$\alpha=-0.76\times 10^{-3}$, confirming Eq.~(\ref{alpha_ll_1}) and we obtain the separation
distance in the twin lead representation $d=2.46$~cm (close to the distance obtained in the 
free space case \cite{full_model_arxiv}, 2.54~cm).

Next using Eq.~(\ref{eps_av}) we obtain $\epsilon_{p}=1.73$, so that $\overline{n}=n_{eq}/\epsilon_{p}=0.93$.
From the cross section analysis we also
obtain the value of the characteristic impedance $Z_0=65.5\Omega$, which is very close to what we obtained in
\cite{full_model_arxiv} for a similar configuration 105.6$\,\Omega$ divided by $n_{eq}=1.613$ (see Eq.~(\ref{Z0})).
This confirms that the electric size of our cross section is small (see discussion at the end of Appendix~A).

We summarise here the parameters used in Eq.~(\ref{rel_P_rad}) for the comparison with simulation:
\begin{equation}
d=2.46\text{cm}\,\,\,n_{eq}=1.613\,\,\,\overline{n}=0.93\,\,\,Z_0=65.5\Omega
\label{params}
\end{equation}

\begin{figure}[!tbh]
\includegraphics[width=9cm]{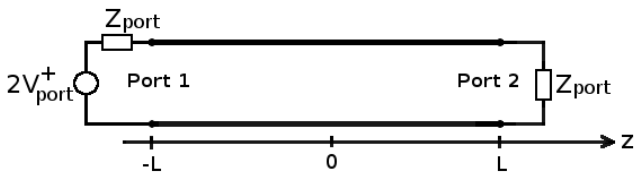}
\caption{Simulation setup for obtaining $2\times 2$ S matrices for different TL lengths.}
\label{two_ports_simulation}
\end{figure}

The simulation setup is shown schematically in Figure~\ref{two_ports_simulation}. The TL is ended at both
sides by lumped ports of characteristic impedance $Z_{port}=50\,\Omega$, but fed only from port 1 by forward
wave voltage $V^+_{port}=1\,V$, so the equivalent Th\'evenin feeding circuit is a generator of $2V^+_{port}$ in series
with a resistance $Z_{port}$.

We obtained from the simulation S matrices defined
for a characteristic impedance $Z_{port}$ at both ports, for different
lengths of the transmission line (similarly to \cite{full_model_arxiv}). By symmetry, the S matrix has the form
\begin{equation}
S=
\left(
\begin{array}{cc}
\Gamma &  \tau  \\
\tau  &    \Gamma 
\end{array}
\right),
\label{S_mat}
\end{equation}
from which one may calculate the ABCD matrix of the TL \cite{pozar, lossless, lossy_MTL, eilat_2014}. We need only the A element
from the ABCD matrix (which is insensitive to the value of $Z_{port}$), as follows:
\begin{equation}
A=\frac{1}{2}\left[\tau+(1-\Gamma^2)/\tau\right]
\label{A_D}
\end{equation}
from which we compute the delay angle of the TL
\begin{equation}
\Theta=\arccos(A)
\label{Theta}
\end{equation}
The real part of $\Theta$ represents the phase accumulated by a forward wave along the TL,
and the imaginary part of $\Theta$ (which is always negative) represents the relative decay of the forward
wave (voltage or current) due to losses (in our case there are only radiation losses) along the TL,
so that $|I^+(L)|=|I^+(-L)|\exp(\text{Im}\{\Theta\})$. Therefore,
the power carried by the forward wave $|P^+(L)|=|P^+(-L)|\exp(2\text{Im}\{\Theta\})$, but for small losses
$|P^+(L)|\simeq |P^+(-L)|(1+2\text{Im}\{\Theta\})$, so that the difference between the input and output values of
$P^+$ (which represent the radiated power $P^+_{rad}$ in Eq.~(\ref{rel_P_rad})), relative to the (average) power $P^+$ carried by the wave is obtained by
\begin{equation}
\frac{P^+_{rad}}{P^+}=-2 \text{Im}\{\Theta\}.
\label{rel_rad_sim}
\end{equation}

\begin{figure}[!tbh]
\includegraphics[width=8cm]{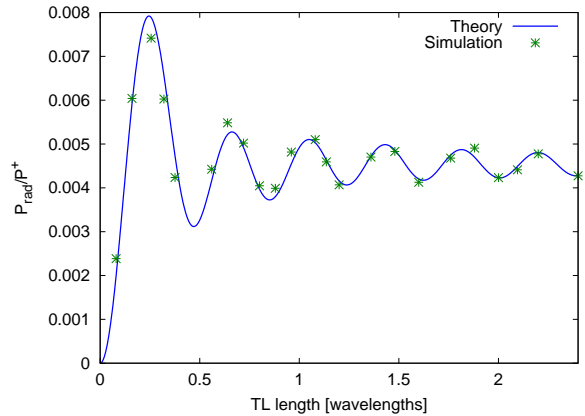}
\caption{Relative radiation losses $P_{rad}/P^+$: comparison between the analytic result in
Eq.~(\ref{rel_P_rad}) and the ANSYS-HFSS simulation result (Eq.~(\ref{rel_rad_sim})) for two parallel cylinders TL.
The horizontal axis is the TL line length in units of wavelengths.}
\label{rad_losses_comparison_dielect}
\end{figure}

In Figure~\ref{rad_losses_comparison_dielect} we compare the analytic result for the relative
power radiated by a forward wave in Eq.~(\ref{rel_P_rad}) with the result in Eq.~(\ref{rel_rad_sim})
obtained from ANSYS-HFSS simulation, at the frequency 240~MHz.
The result shows a very good match between theory and simulation with an average absolute relative
error of 4\%.

For this cross section $\overline{n}$ is 0.93, hence close to 1, so the interference term
(Eq.~(\ref{P_rad_mix})), which scales like $1-\overline{n}^2$ is small. We therefore do not
simulate it for this cross section example, and we shall do it in the next example, as follows.

\subsection{Comparison with ANSYS simulation results - Example 2}

\begin{figure}[!tbh]
\includegraphics[width=8cm]{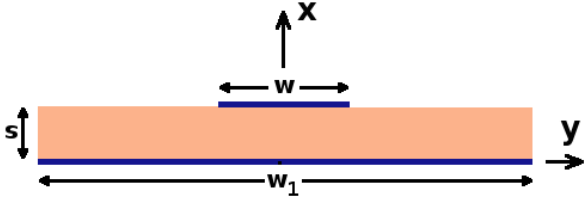}
\caption{Microstrip cross section: the ground conductor (of width $w_1\rightarrow\infty$) is at $x=0$
and the ``plus'' conductor, of width $w=3.4$mm, is located at $x=s=1.52$mm. The conductors are shown
in dark blue and the thickness of the ``plus conductor'' is 17$\mu$m (not mentioned in the figure).
The dielectric insulator (pink) is of relative permittivity $\epsilon_r=3.5$.}
\label{cross_section_ins_ms}
\end{figure}

In this example we use a microstrip cross section shown in Figure~\ref{cross_section_ins_ms}. The width of the
``plus'' conductor is $w=3.4$~mm, the distance between the conductors is $s=1.52$~mm and the relative permittivity
is $\epsilon_r=3.5$. We avoided the conventional
notation $d$ for the distance between the conductors, because $d$ is reserved for the equivalent distance in the twin lead
representation, computed from the cross section analysis (see Appendix~B). However, as expected, for the microstrip case
it comes out that $d$ equals the distance between the conductors, as we shall see in the following cross section analysis.
Using the microstrip formulae \cite{pozar}, we obtain:
\begin{equation}
\epsilon_{eq}=\frac{\epsilon_r+1}{2} + \frac{\epsilon_r-1}{2\sqrt{1+12s/w}}=2.7455
\label{e_eq}
\end{equation}
so that $n_{eq}=\sqrt{\epsilon_{eq}}=1.657$. The characteristic impedance for $w>s$ is given by (see \cite{pozar})
\begin{equation}
Z_0=\frac{\eta_0}{n_{eq} [w/s+1.393+0.667\ln(w/s+1.444)]}=50.55\,\Omega
\label{Z0_formula}
\end{equation}

Because of the ``infinite'' ground conductor in the definition of the microstrip, the theoretical solution
implies 0 fields in the region $x<0$, and of course perpendicular E field and parallel H field on the plane
$x=0^+$. We need to run simulations to determine the fields' structure
in the cross section and to calculate the S parameters for different microstrip lengths, for finding the
radiation losses as function of the TL length, as we did in the previous example. Clearly,
simulations cannot reproduce fields close
to 0 at $x<0$, unless one chooses a very big value for $w_1$, consuming a lot of time and memory. For
values of $w_1$ of the order of $w$ (like $2w$ or $3w$), simulations on the configuration in
Figure~\ref{cross_section_ins_ms} will suffer from significant inaccuracy, not being able to assure
a perpendicular E field and a parallel H field on the plane $x=0^+$.

\begin{figure}[!tbh]
\includegraphics[width=8cm]{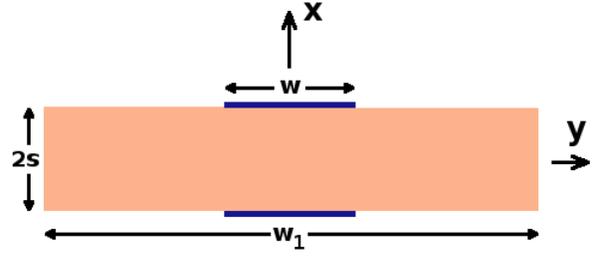}
\caption{Imaged microstrip cross section: the ground conductor at $x=0$ has been eliminated and the
conductor and dielectric at $x>0$ have been imaged to the region $x<0$. For the imaged configuration,
$w_1$ does not need to be very big.}
\label{cross_section_ins_ms_image}
\end{figure}

The method to overcome this is to use an ``imaged'' configuration shown in  
Figure~\ref{cross_section_ins_ms_image}.
The imaged microstrip configuration assures by symmetry perpendicular E field and parallel H field on
the plane $x=0$, independently of the size of $w_1$. It appears that increasing $w_1$ above the value
$2w$ almost does not change the results, hence the choice $w_1=2w$ is very good.

We also remark that for a given forward wave current $I^+$, the imaged configuration carries twice
the power of the original configuration, implying a value of characteristic impedance which is twice
the value in Eq~(\ref{Z0_formula}), i.e.
\begin{equation}
Z_{0\,\text{(imaged)}}=50.55\times 2=101.1\,\Omega
\label{Z0_formula_imaged}
\end{equation}
Also, for a given forward wave current $I^+$, the imaged configuration radiates twice the power of the
original configuration, so that the relative radiated power in Eq.~(\ref{rel_P_rad}) is unchanged.

\begin{figure}[!tbh]
\includegraphics[width=8cm]{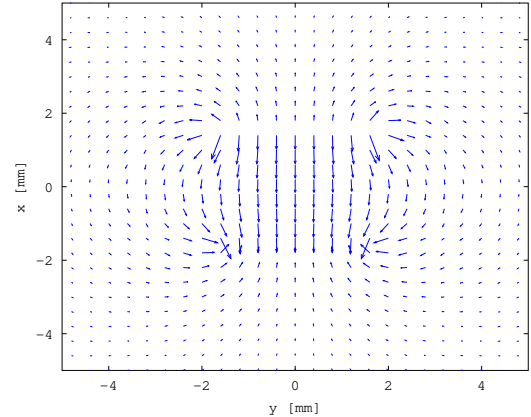}
\caption{Arrow plot of $\mathbf{E}_T$ for the cross section shown in Figure~\ref{cross_section_ins_ms_image}. }
\label{quiver_ms}
\end{figure}

Like in the previous example we perform a cross section analysis, and Figure~\ref{quiver_ms} shows an arrow plot of the transverse electric field $\mathbf{E}_T$. 

From the cross section results, using Eqs.~(\ref{d0_f}), (\ref{alpha_f}) and (\ref{d1_f}), we find that
the weight of the longitudinal polarisation is negligible (as in the previous example) and one obtains
the separation distance in the twin lead representation $d=3.04$~mm. As mentioned previously, we expected
this separation distance to be equal to the distance between the conductors, and indeed it came out $2s$
(see Figure~\ref{cross_section_ins_ms_image}).

From the cross section analysis we also obtain the value of the characteristic impedance $Z_0=99.37\Omega$
(very close to this in Eq.~(\ref{Z0_formula_imaged})) and the equivalent dielectric permittivity $\epsilon_{eq}=2.7$
(very close to this in Eq.~(\ref{e_eq})), so $n_{eq}=\sqrt{\epsilon_{eq}}=1.64$.
Using Eq.~(\ref{eps_av}) we obtain $\epsilon_{p}=2.7$, which equals in this case to $\epsilon_{eq}$,
so that $\overline{n}=1/n_{eq}=0.6$ .

We summarise here the parameters for this cross section:
\begin{equation}
d=3.04\text{mm}\,\,\,n_{eq}=1.657\,\,\,\overline{n}=0.6\,\,\,Z_0=99.37\Omega
\label{params_ms}
\end{equation}

\begin{figure}[!tbh]
\includegraphics[width=8cm]{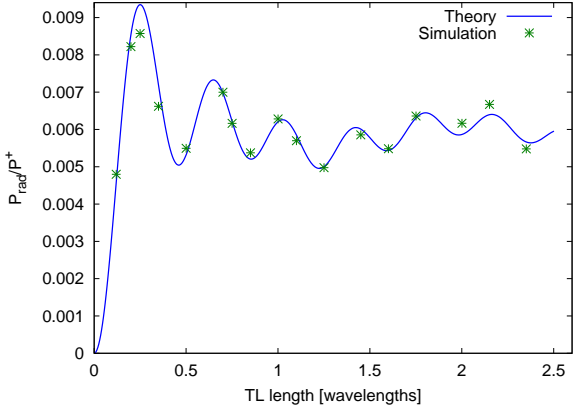}
\caption{Relative radiation losses $P^+_{rad}/P^+$: comparison between the analytic result in
Eq.~(\ref{rel_P_rad}) and the ANSYS-HFSS simulation result (Eq.~(\ref{rel_rad_sim})) for the
(imaged) microstrip TL. The horizontal axis is the TL line length in units of wavelengths.}
\label{rad_losses_comparison_dielect_ms}
\end{figure}

We first compare with ANSYS-HFSS simulation the power radiated by a forward wave, relative to
the power carried by the wave (Eq.~\ref{rel_P_rad}), following the same procedure described in the previous
example, and using the same schematic setup in Figure~\ref{two_ports_simulation}. We obtained S matrices for
different TL length and used Eqs.~(\ref{S_mat})-(\ref{rel_rad_sim}) to elaborate the simulated data.
The comparison is shown in Figure~\ref{rad_losses_comparison_dielect_ms}.
The result shows a very good match between theory and simulation with an average absolute relative
error of 3.2\%.

\begin{figure}[!tbh]
\includegraphics[width=9cm]{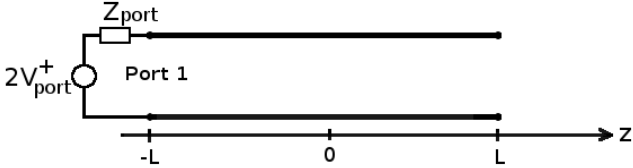}
\caption{Simulation setup for obtaining $S_{11}$ for different TL lengths.}
\label{one_ports_simulation}
\end{figure}

For this cross section $\overline{n}=0.6$ (far from 1), it is therefore expected the interference term
(Eq.~(\ref{P_rad_mix})) to be substantial. We built a simulation setup shown schematically in
Figure~\ref{one_ports_simulation}
to simulate the effect of the interference term. The TL is fed at $z=-L$ by a lumped port of $Z_{port}=50\,\Omega$
with a wave of $V^+_{port}=1\,V$ so the equivalent Th\'evenin feeding circuit is a generator of $2V^+_{port}$ in series
with a resistance $Z_{port}$ (like in the setup from Figure~\ref{two_ports_simulation}), but the right side at $z=L$ is left
open.

Using as first approximation the lossless TL theory, the open end implies $|I^+|=|I^-|$, and the following connection
between the phases of $I^+$ and $I^-$
\begin{equation}
I^+e^{-j\beta L}+I^-e^{j\beta L}=0,
\label{I_plus_I_minus}
\end{equation}
from which
\begin{equation}
\Re\{I^+I^{-*}\}=-|I^+|^2 \cos(2\beta L).
\label{I_plus_I_minus_relation}
\end{equation}
This relation is used in Eq.~(\ref{P_rad_mix}) to calculate the interference term $P_{rad\,\,\,mix}$ contribution in
Eq.~(\ref{P_rad_plus_and_minus}). The value of $|I^+|$ (or $|I^-|$) to be used for the terms $P_{rad}^{\pm}$ in
Eq.~(\ref{P_rad_plus_or_minus}), in terms of $V^+_{port}$ is given by
\begin{equation}
|I^+|=\frac{|V^+_{port}|}{\sqrt{Z_0^2\cos^2(2\beta L)+Z_{port}^2 \sin^2(2\beta L)}}
\label{I_plus_func_V_port}
\end{equation}
Note that the values in Eqs.~(\ref{I_plus_I_minus_relation}) and (\ref{I_plus_func_V_port}) are calculated separately for
each value of $L$, for the comparison with simulation.

\begin{figure}[!tbh]
\includegraphics[width=8cm]{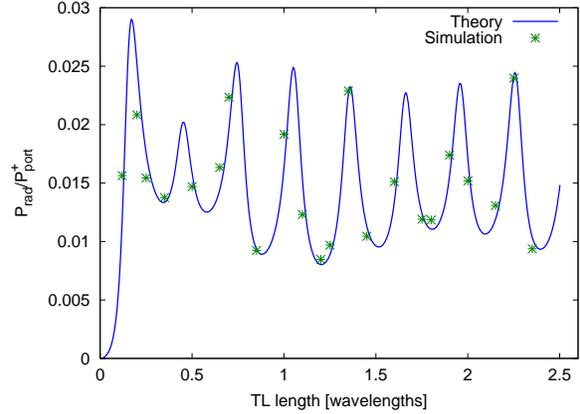}
\caption{Relative radiation losses calculated from Eq.~(\ref{P_rel_open_TL}), compared with $1-|S_{11}|^2$ obtained
from the ANSYS-HFSS simulations for the (imaged) microstrip TL, for different TL lengths. The horizontal axis is the TL
line length in units of wavelengths.}
\label{rel_rad_open_ms}
\end{figure}

By conservation of energy, the power radiated according to Eq.~(\ref{P_rad_plus_and_minus}), must be equal to the power
of the forward wave coming from port 1: $|V^+_{port}|^2/Z_{port}$, multiplied by $1-|S_{11}|^2$. In Figure~\ref{rel_rad_open_ms}
we compare the analytic result from Eq.~(\ref{P_rad_plus_and_minus}) relative to the port power:
\begin{equation}
\frac{P_{rad}}{|V^+_{port}|^2/Z_{port}}
\label{P_rel_open_TL}
\end{equation}
with the values of $1-|S_{11}|^2$ obtained from the ANSYS-HFSS simulation
for different TL lengths.
The result shows a very good match between theory and simulation with an average absolute relative
error of 1.8\%.

\section{Conclusions}

We presented in this work a general algorithm for the analytic calculation of the radiation losses from
transmission lines of two conductors in a dielectric insulator. This work is the generalisation of
\cite{full_model_arxiv}, which deals with TL in free space, and similarly to \cite{full_model_arxiv},
we assumed a small electric cross section, so that the TL carries a Quasi-TEM mode, which behaves similar
to TEM.

We derived the radiation losses of matched TL (carrying a single forward wave) and generalised the result
for non matched TL, carrying any combination of forward and backward waves. Unlike in the free space case
\cite{full_model_arxiv}, the interference between the forward and backward wave have a non zero contribution
to the radiated power, and we successfully validated our analytic results by comparing them with the results of ANSYS-HFSS
simulations for both matched and non matched TL. Also, we compared the matched case with \cite{Nakamura_2006},
but had to neutralise the polarisation currents for the sake of this comparison (given the fact that that
\cite{Nakamura_2006} did not include them in their calculation).

For the specification of the transverse polarisation currents we introduced a new definition of
$\epsilon_{p}$, which basically would mean the usual $\epsilon_{eq}$ if all the polarisation current
elements would be in the same direction (as approximately occurs in a microstrip). However in the
general case, $\epsilon_{p}<\epsilon_{eq}$, where the relation
$\frac{(\epsilon_{p}-1)/\epsilon_{p}}{(\epsilon_{eq}-1)/\epsilon_{eq}}$ represents the average projection of
the polarisation current elements on the main direction of the resultant contribution (set without
loss of generality as the $x$ direction).




\appendices
\renewcommand\thefigure{\thesection.\arabic{figure}}
\setcounter{figure}{0}
\renewcommand\theequation{\thesection.\arabic{equation}}
\setcounter{equation}{0}
\section{Cross section analysis of Quasi-TEM mode}

We perform in this manuscript Quasi-TEM cross sections analyses, to obtain characteristics which affect the radiation
process of dielectric insulated TL. We therefore summarise in this appendix some properties of the cross section solution
for a general case of hybrid TE-TM fields \cite{orfanidis, pozar}. The time dependence is $e^{j\omega t}$, so that
the derivative with respect to time of any variable is a multiplication by $j\omega$.

We call the longitudinal ($z$ directed) fields $E_z$ and $H_z$, and the transverse
($x$ and $y$ components) $\mathbf{E}_T$ and $\mathbf{H}_T$. The transverse ``nabla'' operator is named
$\boldsymbol{\nabla}_T\equiv\widehat{x}\partial_x+\widehat{y}\partial_y$ in Cartesian coordinates. One looks for a
forward wave solution, having the $z$ dependence of the form
\begin{equation}
e^{-j\beta z},
\label{exp_j_beta_z}
\end{equation}
implying
\begin{equation}
\partial_z = -j\beta
\label{partial_z}
\end{equation}
on any variable. This requirement implies the solution of the Helmholtz equations for the longitudinal
fields, and linear relations connecting the transverse fields $\mathbf{E}_T$ and $\mathbf{H}_T$
with $\boldsymbol{\nabla}_T E_z$ and $\boldsymbol{\nabla}_T H_z$, see \cite{orfanidis, pozar}.

For a guided propagation mode, the longitudinal fields are 90 degrees out
of phase relative to the transverse fields, so that any transverse Poynting vector is pure imaginary, which means that
there is a standing wave in the transverse direction and the only net energy is flowing in the longitudinal direction
\cite{orfanidis, pozar}.

It is therefore convenient to scale the phase of the cross section solution so that the transverse fields are real
and the longitudinal fields are pure imaginary, so that:
\begin{equation}
\mathbf{E}_T\equiv \mathbf{E}_{T,R} \,\,\, ; \,\,\, \mathbf{H}_T\equiv \mathbf{H}_{T,R}
\label{E_T_H_T_R}
\end{equation}
and
\begin{equation}
E_z\equiv jE_{z,I} \,\,\, ; \,\,\, H_z\equiv jH_{z,I},
\label{E_z_H_z_I}
\end{equation}
where the ``R'' and ``I'' subscripts indicate real and imaginary parts, respectively. From the cross section
solution one obtains the Quasi-TEM mode propagation wavenumber $\beta$ (see Eqs.~(\ref{exp_j_beta_z}) and
(\ref{partial_z})), from which one defines the effective refraction index of the solution $n_{eq}$ via
\begin{equation}
\beta=n_{eq} k
\label{beta}
\end{equation}
where
\begin{equation}
k=\omega/c
\label{k}
\end{equation}
is the free space wavenumber, $c$ being the free space speed of light in vacuum. This equivalent refraction index
satisfies $1\le n_{eq} \le n$, according to ``how much''
fields are in the dielectric and ``how much'' in the surrounding air, where $n=\sqrt{\epsilon_r}$ is the refraction
index of the dielectric material.

To be mentioned that such mode is called Quasi-TEM, because it behaves close to TEM, in the sense that
the transverse fields $\mathbf{E}_T$ and $\mathbf{H}_T$ are dominant relative to the longitudinal fields
$E_z$ and $H_z$. This may be symbolically written as
\begin{equation}
E_T, (\eta_0/n_{eq})H_T \gg E_z, (\eta_0/n_{eq})H_z,
\label{transverse_dominant}
\end{equation}
in the averaging sense ($\eta_0=377\Omega$ is the free space impedance). As smaller the electrical size
of the cross section, the above condition is more accurate.

Now examining the surface free current continuity on one of the conductors in Figure~\ref{config}
(say the ``plus'' conductor having the contour $c_1$), we obtain
\begin{equation}
\partial_t \rho_s + \partial_z J_{s\,z} +\partial_{c1} J_{s\,c1}=0
\label{surface_free_current_continuity}
\end{equation}
where $\rho_s$ is the free surface charge and $J_{s\,z}$ and $J_{s\,c1}$ are the longitudinal
and transverse components of the free surface current. Using $\partial_t=j\omega$ and
$\partial_z = -j\beta$ (Eq.~(\ref{partial_z})), we get
\begin{equation}
j\omega\rho_s - j\beta J_{s\,z} +\partial_{c1} J_{s\,c1}=0
\label{surface_free_current_continuity1}
\end{equation}
Clearly the free transverse surface current $J_{s\,c1}$ which is proportional to $H_z$ is negligible relative
to longitudinal surface current $J_{s\,z}$ which is proportional to $\mathbf{H}_T$
(see Eq.~(\ref{transverse_dominant})). But for now we do not need this condition, because the last term in
Eq.~(\ref{surface_free_current_continuity1}) vanishes after integrating around the contour
$\oint_{c1}$, obtaining
\begin{equation}
j\omega\rho_l - j\beta I^+ =0,
\label{surface_free_current_continuity2}
\end{equation}
where $\rho_l$ is the charge per longitudinal length unit. We shall use $I^+$, $V^+$ for the forward current
and voltage waves, respectively (we deal with a forward wave, so in principle, all quantities should have a
``+'' superscript, but it would be too cumbersome). Here we use the capacitance per
length unit defined via $\rho_l=CV^+$:
\begin{equation}
\omega CV^+ = \beta I^+,
\label{surface_free_current_continuity3}
\end{equation}
and for this we do need condition (\ref{transverse_dominant}), because the potential
difference between the conductors $V^+$ has a meaning only if its value $\int \mathbf{E}_T\cdot \mathbf{dl}$
is independent of the integration trajectory, and this is strictly correct only if $H_z=0$.

Supposing condition (\ref{transverse_dominant}) is satisfied and using Eqs.~(\ref{beta}), (\ref{k}), and the definition
of the characteristic impedance $Z_0\equiv V^+/I^+$, we obtain
\begin{equation}
Z_0  = \frac{n_{eq}}{cC},
\label{Z0_1}
\end{equation}
and comparing it to the ``telegraph model'' definition $Z_0\equiv\sqrt{L/C}$, $L$
being the inductance per length unit, one obtains
\begin{equation}
\sqrt{LC}=n_{eq}/c
\label{LC}
\end{equation}
Dealing with dielectric materials, the inductance per unit length $L$ is the same as the free space
inductance, we therefore conclude from Eq.~(\ref{LC}) that $C$ is proportional to
$n_{eq}^2=\epsilon_{eq}$, so that
\begin{equation}
C=\epsilon_{eq} C_{\text{free space}}.
\label{C}
\end{equation}
From Eqs.~(\ref{Z0_1}) and (\ref{C}) it results that $Z_0$ is inverse proportional to
$n_{eq}$, so that
\begin{equation}
Z_0=Z_{0\,\text{free space}}/n_{eq}
\label{Z0}
\end{equation}
We may understand from this analysis that the relation between $\beta$ and $k$, namely $n_{eq}$
describes in the DC limit the connection between the capacitance per unit length (or the
characteristic impedance) with dielectric and the parallel value in free space according to
Eqs.~(\ref{C}) and (\ref{Z0}). Therefore the relation between $\beta$ and $k$ keeps linear
as long as the electric size of the cross section is small, and may deviate from this relation
for higher frequencies.

\renewcommand\thefigure{\thesection.\arabic{figure}}
\setcounter{figure}{0}
\renewcommand\theequation{\thesection.\arabic{equation}}
\setcounter{equation}{0}
\section{Far vector potential of separated two-conductors transmission line in twin lead representation}

In this appendix we calculate the far magnetic vector potential for a general TL insulated in a dielectric
material of relative dielectric permittivity $\epsilon_r$ as shown in Figure~\ref{config}.

In spite of dealing with dielectric insulators on has to use the regular free space Lorenz gauge \cite{Krey},
in the frequency domain
\begin{equation}
\boldsymbol{\nabla}\cdot\mathbf{A}+\frac{j\omega V}{c^2}=0,
\label{lorenz_gauge}
\end{equation}
obtaining the following wave equations for the magnetic vector potential $\mathbf{A}$ and the scalar
potential $V$ \cite{Krey}:
\begin{equation}
\left(\nabla^2+k^2\right)\mathbf{A}=-\mu_0\mathbf{J}_{\text{eff}}
\label{A_wave_eq}
\end{equation}
\begin{equation}
\left(\nabla^2+k^2\right)V=-\rho_{\text{eff}}/\epsilon_0.
\label{V_wave_eq}
\end{equation}
where
\begin{equation}
\mathbf{J}_{\text{eff}}=\mathbf{J}+j\omega\mathbf{P}+\boldsymbol{\nabla}\times\mathbf{M},
\label{J_eff}
\end{equation}
\begin{equation}
\rho_{\text{eff}}=\rho-\boldsymbol{\nabla}\cdot\mathbf{P},
\label{rho_eff}
\end{equation}
$\mathbf{J}$ and $\rho$ being the free current and charge densities, respectively,
\begin{equation}
\mathbf{P}=\epsilon_0(\epsilon_r-1)\mathbf{E}
\label{P}
\end{equation}
is the electric polarisation field and
\begin{equation}
\mathbf{M}=0
\label{M}
\end{equation}
is the magnetisation field which is 0 in our case.

We remark that $\mathbf{J}_{\text{eff}}$ and $\rho_{\text{eff}}$ satisfy the continuity equation
\begin{equation}
\boldsymbol{\nabla}\cdot\mathbf{J}_{\text{eff}}+j\omega \rho_{\text{eff}}=0,
\label{continuity_equation}
\end{equation}
(which holds separately for the free and polarisation charges/currents). We see that $\rho_{\text{eff}}$ can be calculated from $\mathbf{J}_{\text{eff}}$, the same as
$V$ can be calculated from $\mathbf{A}$ using (\ref{lorenz_gauge}). This means that we have
to solve only Eq.~(\ref{A_wave_eq}), as usually done in radiation problems
\cite{orfanidis, ramo, jordan, balanis, stutzman_thiele}. Its formal solution, for a TL
from $-L$ to $L$ is the convolution integral
\begin{equation}
\mathbf{A}(x,y,z)=\mu_0\int_{-L}^{L} dz' \iint\limits_{\substack{\text{TL cross} \\ \text{section}}} dx' dy' \mathbf{J}_{\text{eff}}(x',y',z')\, G(R)
\label{vec_A_basic_f}
\end{equation}
where
\begin{equation}
G(s)=\frac{e^{-jks}}{4\pi s}
\label{Green}
\end{equation}
is the 3D Green's function and
\begin{equation}
R=\sqrt{(x-x')^2+(y-y')^2+(z-z')^2}.
\label{R}
\end{equation}
We remark that (\ref{vec_A_basic_f}) is the potential vector due to the currents along the TL, and the
contribution of the termination (source and load) currents \cite{full_model_arxiv} are calculated at the
end of this appendix.

According to (\ref{exp_j_beta_z}), we use
\begin{equation}
\mathbf{J}_{\text{eff}}(x',y',z')=\mathbf{J}_{\text{eff}}(x',y')e^{-j\beta z},
\label{J_eff_x_y_z}
\end{equation}
and approximating $R$ in (\ref{R}) in far field in spherical coordinates to
\begin{equation}
R=r-(x'\cos\varphi+y'\sin\varphi)\sin\theta-z'\cos\theta
\label{R_far_spherical}
\end{equation}
we rewrite $\mathbf{A}$ in the far field
\begin{align}
\mathbf{A}=&\mu_0 G(r)\int_{-L}^{L} dz' e^{jkz'(\cos\theta-n_{eq})} \notag \\
& \iint\limits_{\substack{\text{TL cross} \\ \text{section}}} dx' dy' \mathbf{J}_{\text{eff}}(x',y')e^{jk\sin\theta [x'\cos\varphi+y'\sin\varphi]}.
\label{vec_A_basic_1_f}
\end{align}
At this point the $z'$ integral can be separated from the cross section integral. Integrating on $z'$
we obtain
\begin{equation}
\mathbf{A}=\mu_0 G(r)2L\sinc[kL(\cos\theta-n_{eq})] \mathbf{Q}(\theta,\varphi)
\label{vec_A_basic_2_f}
\end{equation}
where $\sinc(x)\equiv\sin x/x$ and $\mathbf{Q}(\theta,\varphi)$ is
\begin{equation}
\mathbf{Q}= \iint\limits_{\substack{\text{TL cross} \\ \text{section}}} dx' dy' \mathbf{J}_{\text{eff}}(x',y')e^{jk\sin\theta [x'\cos\varphi+y'\sin\varphi]}.
\label{Q}
\end{equation}
so that the direction of $\mathbf{A}$ is according to the direction of $\mathbf{Q}$.

Considering the higher modes to be in deep cutoff, so that $kx',ky'\ll 1$ (small electric cross section), and
defining the radial cross section unit vector
\begin{equation}
\boldsymbol{\widehat{\rho}}(\varphi)=\mathbf{\widehat{x}}\cos\varphi+\mathbf{\widehat{y}}\sin\varphi
\label{widehat_rho}
\end{equation}
and the radial cross section integration variables vector
\begin{equation}
\boldsymbol{\rho}'\equiv x'\mathbf{\widehat{x}}+y'\mathbf{\widehat{y}}.
\label{rho_prime_vec}
\end{equation}
we may rewrite Eq.~(\ref{Q}) as
\begin{equation}
\mathbf{Q}=\iint\limits_{\substack{\text{TL cross} \\ \text{section}}} dx' dy' \mathbf{J}_{\text{eff}} [1+jk\sin\theta\boldsymbol{\widehat{\rho}}(\varphi)\cdot\boldsymbol{\rho}'].
\label{Q1}
\end{equation}
The strategy to calculate $\mathbf{Q}$ is as follows: for components of $\mathbf{J}_{\text{eff}}$ on which the integral
$dx'dy'$ vanishes
over the TL cross section, we perform the integral of the component multiplied by 
$jk\sin\theta\boldsymbol{\widehat{\rho}}(\varphi)\cdot\boldsymbol{\rho}'$, as follows
\begin{equation}
jk\sin\theta \iint\limits_{\substack{\text{TL cross} \\ \text{section}}} dx' dy' \mathbf{J}_{\text{eff}}\, [\boldsymbol{\widehat{\rho}}(\varphi)\cdot\boldsymbol{\rho}'],
\label{Q2}
\end{equation}
while for components of $\mathbf{J}_{\text{eff}}$ on which the integral $dx'dy'$ is not 0,
we neglect $k\sin\theta\boldsymbol{\widehat{\rho}}(\varphi)\cdot\boldsymbol{\rho}'$ relative to 1,
as follows
\begin{equation}
\iint\limits_{\substack{\text{TL cross} \\ \text{section}}} dx' dy' \mathbf{J}_{\text{eff}}
\label{Q3}
\end{equation}

For the free space TL \cite{full_model_arxiv} we had to deal only with the longitudinal ($z$ component) of $\mathbf{A}$ in
what concerns the TL currents contribution (which were free surface currents in the $z$ direction), and we had
transverse components ($x$ or $y$) of $\mathbf{A}$ only from the terminations of the TL. In the current case we note that
$\mathbf{J}_{\text{eff}}$ contains both free longitudinal surface currents and polarisation currents contributions
(which have both longitudinal and transverse components). We therefore deal first with the longitudinal component
of $\mathbf{Q}$, which is written as:
\begin{equation}
Q_z=Q_{z\,\,\text{free}}+Q_{z\,\,\text{pol}},
\label{Q_z}
\end{equation}
where $Q_{z\,\,\text{free}}$ is the contribution of the longitudinal free surface currents $K_z$ and hence is similar
to \cite{full_model_arxiv} (so that the solution to Eq.~(\ref{Q1}) has the form (\ref{Q2})):
\begin{equation}
Q_{z\,\,\text{free}}=jk\sin\theta\boldsymbol{\widehat{\rho}}(\varphi)\cdot\oint dc\, K_z(c)\boldsymbol{\rho}'(c),
\label{Q_z_free}
\end{equation}
where $c$ is the contour parameter around the perfect conductors (i.e. $c_1$ and $c_2$, see Figure~\ref{config}).
Separating the contours and noting that we deal with a differential mode for which the currents in the conductors
are equal but with opposite signs, and using $K_z=H_{T\,\parallel}$ (i.e. the component of $\mathbf{H}_T$ parallel
to the conductors) one obtains:
\begin{equation}
Q_{z\,\,\text{free}}=jk\sin\theta  I^+ \mathbf{d}_0\cdot\boldsymbol{\widehat{\rho}}
\label{Q_z_free1}
\end{equation}
the forward current $I^+$ is
\begin{equation}
I^+=\oint dc_1 H_{T\,\parallel}(c_1)=-\oint dc_2 H_{T\,\parallel}(c_2),
\label{I_plus_f}
\end{equation}
\begin{equation}
\mathbf{d}_0\equiv \oint dc H_{T\,\parallel}(c)\boldsymbol{\rho}'(c)/I^+,
\label{vec_d0_f}
\end{equation}
The vector $\mathbf{d}_0$ represents the vector distance pointing from the ``negative'' conductor
to the ``positive'' conductor in the twin lead equivalent (see Figure~\ref{twin_lead}). As in \cite{full_model_arxiv}
it is convenient to redefine the $x$ axis to be aligned with $\mathbf{d}_0$, so that $d_{0x}=d_0$ and $d_{0y}=0$,
so that (\ref{Q_z_free1}) simplifies to
\begin{equation}
Q_{z\,\,\text{free}}=jk\sin\theta  I^+ d_0 \cos\varphi
\label{Q_z_free2}
\end{equation}
and the expression for the distance $d_0$ simplifies to
\begin{equation}
d_0\equiv \oint dc H_{T\,\parallel}(c)x'(c)/I^+,
\label{d0_f}
\end{equation}
For $Q_{z\,\,\text{pol}}$ in Eq.~(\ref{Q_z}) we use the longitudinal polarisation current density
$j\omega\epsilon_0(\epsilon_r-1)E_z$ (see Figure~\ref{config}), which according to (\ref{E_z_H_z_I}) can be written
as $-\omega\epsilon_0(\epsilon_r-1)E_{z,I}$, and the integration is only on the dielectric region.
Clearly, $E_z$ being a solution of the Helmholtz equation, the integral on the TL cross section
vanishes, and for the twin lead representation, we obtain
\begin{equation}
Q_{z\,\,\text{pol}}=\iint\limits_{\substack{\text{dielectric} \\ \text{region}}} dx' dy' [-\omega\epsilon_0(\epsilon_r-1)]E_{z,I} jk\sin\theta x'\cos\varphi,
\label{Q_z_pol}
\end{equation}
which can be written as
\begin{equation}
Q_{z\,\,\text{pol}}=jk\sin\theta \alpha I^+ d_1 \cos\varphi
\label{Q_z_pol1}
\end{equation}
where
\begin{equation}
\alpha\equiv -\omega\epsilon_0(\epsilon_r-1)\iint\limits_{\substack{\text{dielectric} \\ \text{region}}} dx' dy' E_{z,I}\mathcal{H}(E_{z,I})/I^+,
\label{alpha_f}
\end{equation}
and $\mathcal{H}$ represents the Heaviside step function, limiting the integral to the regions in which $E_{z,I}>0$ and
\begin{equation}
d_1\equiv \frac{\iint\limits_{\substack{\text{dielectric} \\ \text{region}}} dx' dy' E_{z,I} x'}{\iint\limits_{\substack{\text{dielectric} \\ \text{region}}} dx' dy' E_{z,I}\mathcal{H}(E_{z,I})}.
\label{d1_f}
\end{equation}
Clearly, $\alpha$ being the ratio between something proportional to $E_z$ and $I^+$, which is proportional to
$H_T$, satisfies 
\begin{equation}
|\alpha|\ll 1
\label{alpha_ll_1}
\end{equation}
(see Eq.~(\ref{transverse_dominant})). Using the above definitions, we sum Eqs.~(\ref{Q_z_free2}) and (\ref{Q_z_pol1}) to obtain the total $z$ component of
$\mathbf{Q}$
\begin{equation}
Q_z=jk\sin\theta I^+ d\cos\varphi
\label{Q_z1}
\end{equation}
where
\begin{equation}
d=d_0+\alpha d_1\simeq d_0
\label{d}
\end{equation}
is the separation distance between the conductors in the twin lead representation, analogous to what we obtained in
\cite{full_model_arxiv}.
For the case of conductors in dielectric insulator this vector separation has two contributions: $d_0$ is due to
the free currents,
$d_1$ due to the polarisation currents and $\alpha$ is the ``weight'' of the longitudinal polarisation
contribution, but as explained above, this weight is typically small (see Eq.~(\ref{alpha_ll_1})).

Now we calculate the transverse component of $\mathbf{Q}$, which under the twin lead representation
simplifies to $\mathbf{\widehat{x}}Q_x$ (see Figure~\ref{twin_lead}), where $Q_x$ is due to the transverse polarisation
currents. We therefore use for $J_\text{eff\, x}$ the $x$ directed polarisation current density $J_{p\,x}$
(and the solution to Eq.~(\ref{Q1}) has the form (\ref{Q3})):
\begin{equation}
Q_x=\iint\limits_{\substack{\text{dielectric} \\ \text{region}}} dx' dy' J_{p\,x},
\label{Q_x}
\end{equation}
where
\begin{equation}
J_{p\,x}=j\omega P_x=j\omega\epsilon_0(\epsilon_r-1)E_x,
\label{J_p1}
\end{equation}
and $E_x$ is the $x$ component of $\mathbf{E}_T$.

It is useful to describe the transverse polarisation current in the twin lead representation as a
$-\mathbf{\widehat{x}}$ directed surface current on the surface $y=0$ (see Figure~\ref{twin_lead}). The
polarisation (surface) current is proportional to the displacement surface
current:
\begin{equation}
J_{sd\,x}=-j\omega C V^+ ,
\label{J_s_d}
\end{equation}
where the minus is due to the fact that the current is in the $-\mathbf{\widehat{x}}$ direction. Using relations
(\ref{beta}) and (\ref{Z0_1}), this can be written
\begin{equation}
J_{sd\,x}=-j\beta I^+
\label{J_s_d1}
\end{equation}
Inside a uniform dielectric material the polarisation current is $(\epsilon_r-1)/\epsilon_r$ times the
displacement current, but having part of the fields is in air, it looks like one has to use
$(\epsilon_{eq}-1)/\epsilon_{eq}$ times the displacement current. As shown in Appendix~C, due to the
fact that parts of the polarisation current elements are in the perpendicular ($y$) direction,
if one wants to use the same value for $d$ as defined in Eq.~(\ref{d}), one needs to use a value
for the relative dielectric permittivity in general smaller than $\epsilon_{eq}$, which we name
$\epsilon_{p}$ (the subscript ``p'' stands for polarisation), for expressing the polarisation surface current:
\begin{equation}
J_{sp\,x}=\frac{\epsilon_{p}-1}{\epsilon_{p}}J_{sd\,x},
\label{J_s_p}
\end{equation}
and being a surface current on $y=0$, one gets the polarisation current density
\begin{equation}
J_{p\,x}=J_{sp\,x}\delta(y)
\label{J_p}
\end{equation}
The value of $\epsilon_{p}$ satisfies:
\begin{equation}
1\le\epsilon_{p}\le\epsilon_{eq},
\label{epsilon_av_epsilon_eq}
\end{equation}
and Appendix~C is dedicated to explain this connection, the physical meaning of $\epsilon_{p}$ and its
relation to $\epsilon_{eq}$.

Using (\ref{J_p}) in (\ref{Q_x}), the integral $dx'$ is carried out
from 0 to $d$, while the $dy'$ integral yield 1, because of the delta function, obtaining
\begin{equation}
Q_x=\int_0^d dx' J_{sp\,x}=d J_{sp\,x}=-j\beta I^+ d \frac{\epsilon_{p}-1}{\epsilon_{p}}.
\label{Q_x1}
\end{equation}
By comparing (\ref{Q_x1}) with (\ref{Q_x}) and using (\ref{J_p1}), one obtains an equation to calculate $\epsilon_{p}$
from the numerical cross section solution, as follows:
\begin{equation}
\frac{\epsilon_{p}-1}{\epsilon_{p}}=-\frac{\epsilon_r-1}{n_{eq}\eta_0 I^+ d}\iint\limits_{\substack{\text{dielectric} \\ \text{region}}} dx' dy' E_x,
\label{eps_av}
\end{equation}
where the RHS is positive, because the phase of $E_x$ is scaled to point mainly from the ``positive'' to the
``negative'' conductor, i.e. it is mainly negative.

At this point we summarise the results for the vector potential components contributed by the currents along
the TL. From (\ref{vec_A_basic_2_f}) and (\ref{Q_z1}) we obtain
\begin{equation}
A_z=\mu_0 G(r)2L\sinc[kL(\cos\theta-n_{eq})] jk\sin\theta I^+ d\cos\varphi,
\label{A_z}
\end{equation}
and from (\ref{vec_A_basic_2_f}) and (\ref{Q_x1}) we get the $x$ directed contribution. Given this
contribution is from transverse polarisation, we name it $A_{x\,\text{pol}}$:
\begin{align}
A_{x\,\text{pol}}=&\mu_0 G(r)2L\sinc[kL(\cos\theta-n_{eq})] \notag \\
               &(-j\beta I^+) d \frac{\epsilon_{p}-1}{\epsilon_{p}}.
\label{A_x_pol}
\end{align}
It is worthwhile to mention that any representation that keeps the value of $d(\epsilon_{p}-1)/\epsilon_{p}$
correct, for example replacing $\epsilon_{p}$ by $\epsilon_{eq}$, but accordingly use a smaller value of $d$ for the
transverse polarisation would be a completely equivalent representation, but we chose to keep the same value
of $d$ for representing the free currents and the polarisation currents (see discussion in Appendix~C).

Now we calculate the contribution of the termination currents to the potential vector. The twin lead geometry
allows us to use a simple model for the termination currents, which are in the $x$ direction (see Figure~\ref{twin_lead}),
and their values are $\pm I^+e^{\pm j\beta L}$ at the locations $\mp L$ respectively. They result in
\begin{equation}
A_{x\, 1,2}=\pm\mu_0I^+\int_{-d/2}^{d/2}dx' e^{\pm j\beta L}G(R_{1,2})
\label{A_x12_f}
\end{equation}
where the indices 1,2 denote the contributions from the termination currents at $\mp L$, respectively,
(see Figure~\ref{twin_lead}). The distances $R_{1,2}$ of the far observer from the terminations 
may be expressed in spherical coordinates, as follows:
\begin{equation}
R_{1,2}\simeq r - z_{1,2}\cos\theta - x'\sin\theta\cos\varphi,
\label{R_12_f}
\end{equation}
The integral (\ref{A_x12_f}) is carried out for $kd\ll 1$, using (\ref{beta}), results in
\begin{equation}
A_{x\, 1,2}=\pm\mu_0I^+d G(r)e^{\mp jk L(\cos\theta-n_{eq})}.
\label{A_x12_1_f}
\end{equation}
The two contributions sum to $A_{x\, 1}+A_{x\, 2}$:
\begin{equation}
A_{x\, \text{free}}=\mu_0G(r)I^+d (-2j)\sin[kL(\cos\theta-n_{eq})],
\label{A_x_free}
\end{equation}
and we call it $A_{x\, \text{free}}$, because the termination currents are free currents. The total transverse
$x$ directed potential vector is obtained by summing (\ref{A_x_pol}) with (\ref{A_x_free}):
\begin{align}
A_x=&\mu_0G(r)I^+d 2L\sinc[kL(\cos\theta-n_{eq})](-jk) \notag \\
    &(\cos\theta-n_{eq}/\epsilon_{p})
\label{A_x}
\end{align}

\renewcommand\thefigure{\thesection.\arabic{figure}}
\setcounter{figure}{0}
\renewcommand\theequation{\thesection.\arabic{equation}}
\setcounter{equation}{0}
\section{The physical meaning of $\epsilon_{eq}$, $\epsilon_{p}$ and the connection between them}

We used in this work a new quantity called $\epsilon_{p}$ for the purpose of defining the
contribution of the transverse polarisation currents. This appendix is dedicated to explain
the physical meaning of $\epsilon_{p}$ and show the connection between it and $\epsilon_{eq}$.

First, it should be mentioned that it is not the polarisation
current per se which affects the radiation, but rather the polarisation current {\it element}, i.e.
$d J_{sp\,x}$ - see Eq.~(\ref{Q_x1}). Looking at Eq.~(\ref{eps_av}), it is clear that the value
of $d(\epsilon_{p}-1)/\epsilon_{p}$ is only a function of the cross section geometry
(while $d$ is something that we defined in Eq.~(\ref{d}) to formulate the twin lead model).

This means that the only requirement to obtain a correct expression for the radiation is to use
the correct value of $d(\epsilon_{p}-1)/\epsilon_{p}$, and we had the freedom to replace
$\epsilon_{p}$ by $\epsilon_{eq}$ and determine accordingly a new value for the equivalent
distance for the polarisation contribution and call it $d_p$ for example. This would imply
\begin{equation}
d\frac{\epsilon_{p}-1}{\epsilon_{p}}=d_p\frac{\epsilon_{eq}-1}{\epsilon_{eq}},
\label{alternative}
\end{equation}
so the usage of $\epsilon_{eq}$ and $d_p$ 
gives a completely equivalent formulation, leading to the same result.
The only reason we did not choose it is aesthetic: we simply preferred a uniform
twin lead model for both free currents and polarisation, having the same effective
separation distance $d$.

\begin{figure}[!tbh]
\includegraphics[width=8cm]{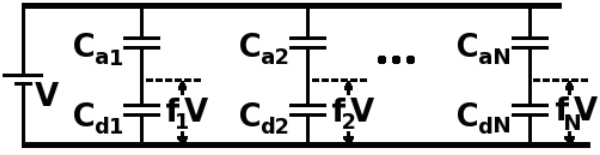}
\caption{The space between the conductors is modelled by capacitors in parallel $C_1$, $C_2$, ... $C_N$, each
 one representing a slice around an electric field line (shown in Figure~\ref{field_lines}). Each electric field line
 may pass part of its
trajectory through the air, and the other part through the dielectric, hence each capacitor $C_i$ consists of two
capacitors in series: $C_{a\, i}$ and $C_{d\, i}$, so that $C_i^{-1}=C_{a\, i}^{-1}+C_{d\, i}^{-1}$. The value $f_i$ is the
fraction of voltage on the dielectric, for the field line defining capacitor $C_i$ (see Figure~\ref{field_lines}).}
\label{capacitors}
\end{figure}

We emphasise this point, because we need a consistent definition for the separation distance to analyse the polarisation
current elements in an equivalent circuit that we develop here (shown in Figure~\ref{capacitors}).
For this purpose we use the same value $d$ used along the whole paper, i.e. this one given in Eq.~(\ref{d}).

\begin{figure}[!tbh]
\includegraphics[width=8cm]{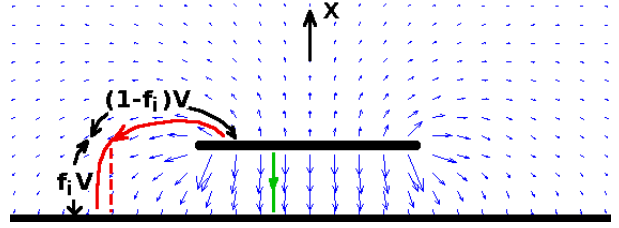}
\caption{The transverse electric field in a microstrip shown by the blue arrows. The total voltage on the microstrip is $V$.
In general, field lines are partly in the air and partly in the dielectric, and we define the voltage on the air
$(1-f_i)V$ and the voltage on the dielectric $f_iV$ for a given electric field line $i$.
This is evidently shown on the ``red'' line which is partly in the air and partly in the dielectric.
The ``green'' line passes only through the dielectric, it is therefore a special case of the above with $f_i=1$.
The dashed ``red'' line shows the
projection of the continuous line (inside the dielectric) on the $x$ direction, and we define the relation between
the (dashed) projection length and the continuous line $g_i$ for a given electric field line $i$. For the red line
$g_i$ is close but not equal to 1, but for the ``green'' line, being parallel to the $x$ direction, $g_i=1$.}
\label{field_lines}
\end{figure}

We consider the capacitors in Figure~\ref{capacitors} as parallel plates capacitors, as follows
\begin{equation}
C_{a\,i} \equiv \frac{\epsilon_0 A_0}{d_{a\,i}}\,\,\,;\,\,\,C_{d\,i} \equiv \frac{\epsilon_0 \epsilon_r A_0}{d_{d\,i}},
\label{C_air_and_d}
\end{equation}
where $A_0$ is a {\it fixed} effective area (more accurately perpendicular length), $d_{a\,i}$ and $d_{d\,i}$ are
the effective separation distances of the air and dielectric parts, respectively, and we require their sum to be
the total effective separation distance $d$ mentioned before:
\begin{equation}
d_{a\,i}+d_{d\,i}=d.
\label{d_a_and_d_d}
\end{equation}
From Eqs.~(\ref{C_air_and_d}) and (\ref{d_a_and_d_d}) it is easy to show that $d_{a\,i}$ and $d_{d\,i}$ come out
\begin{equation}
d_{a\,i} = d \frac{1-f_i}{1-f_i+\epsilon_r f_i}\,\,\,;\,\,\,d_{d\,i} =d \frac{\epsilon_r f_i}{1-f_i+\epsilon_r f_i},
\label{d_air_and_d}
\end{equation}
and the capacitor $C_i$ may be written as
\begin{equation}
C_i \equiv \frac{\epsilon_0 A_0}{d} [1-f_i+\epsilon_r f_i]
\label{C_i}
\end{equation}
The field lines describing the capacitors $C_i$ in the equivalent circuit in Figure~\ref{capacitors} are shown in
Figure~\ref{field_lines} for a microstrip.
In general each field line is partly in the air and partly in the dielectric, so that the voltage on the air is
$\int_{\substack{\text{air} \\ \text{part}}} \mathbf{E}_T\cdot \mathbf{dl}$ and the voltage on the dielectric is
$\int_{\substack{\text{dielectric} \\ \text{part}}} \mathbf{E}_T\cdot \mathbf{dl}$. According to this, we defined $f_i$
the fraction of voltage on the dielectric, as evidently shown on the ``red'' line in Figure~\ref{field_lines}.
The ``green'' line in Figure~\ref{field_lines} passes only through the dielectric, it is therefore a special case
of the above with $f_i=1$. The dashed ``red'' line shows the projection of the continuous line
(inside the dielectric) on the $x$ direction, and the relation between the projected line and the original line
is called $g_i$ for the electric field line $i$. For the red line, $g_i$ is not equal, but close to 1. For the green line,
being in the $x$ direction, $g_i=1$. The projection is discussed further on in context with the polarisation currents.

Given the number of capacitors is $N$, the total capacitance per longitudinal unit length $C$ is equal the sum on
$i$ of all the capacitances $C_i$ in Eq.~(\ref{C_i})
\begin{equation}
C=\sum_{i=1}^NC_i = \frac{\epsilon_0 A_0}{d} \sum_{i=1}^N [1-f_i+\epsilon_r f_i].
\label{total_C}
\end{equation}
The free space capacitance is obtained by setting all $f_i=0$: $C_{\text{free space}}=\frac{N\epsilon_0 A_0}{d}$, and
the value of $\epsilon_{eq}$ is calculated from Eq.~(\ref{C}), obtaining
\begin{equation}
\epsilon_{eq}= \frac{1}{N}\sum_{i=1}^N [1-f_i+\epsilon_r f_i]=1+(\epsilon_r-1)\frac{1}{N}\sum_{i=1}^N f_i
\label{epsilon_eq}
\end{equation}
or in a more suggestive form:
\begin{equation}
\frac{\epsilon_{eq}-1}{\epsilon_r-1}=\langle f \rangle,
\label{epsilon_eq_1}
\end{equation}
where $\langle f \rangle$ is the average fraction of voltage on the dielectric. Given that electric field is
proportional to voltage, and polarisation vector is proportional to $\epsilon_r-1$ times electric field
(see Eq.~(\ref{P})), suggests that $\epsilon_{eq}-1$ indicates on the average polarisation vector. We remark that
using Eqs.~(\ref{total_C}) and (\ref{epsilon_eq}), one can write the total capacitance as
\begin{equation}
C=\frac{\epsilon_0 \epsilon_{eq}(NA_0)}{d}
\label{total_C_1}
\end{equation}
so that it is represented by a parallel plates capacitor of relative dielectric permittivity $\epsilon_{eq}$, distance
$d$ between the plates and area (or rather perpendicular length) $NA_0$.

Now we calculate the polarisation current element. On a given capacitor, the displacement current is $I_D=j\omega C V$,
which is also the (AC) current passing through the capacitor. We remark that this total current is a cross section
integral on a current density vector in the space between the plates, and this vector may have different directions
in different locations. The total effect on the radiation comes from the equivalent polarisation current {\it element} vector
contribution (we chose the $x$ axis in this direction - see Appendix~B). We therefore need the $x$ projection of the
polarisation current element vector (see projection factor $g$ - dashed red line in Figure~\ref{field_lines}) - we shall
call it $Q_P$. It is obtained by multiplying $I_D$ by $d(\epsilon_{p}-1)/\epsilon_{p}$, where $\epsilon_{p}$ already includes
effect of the projection, as explained at the beginning of this appendix. Considering the parallel plate capacitor of our model
in Eq.~(\ref{total_C_1}), we have
\begin{equation}
Q_P=d\frac{\epsilon_r-1}{\epsilon_r} I_D=j\omega V\epsilon_0\epsilon_{eq} NA_0 (\epsilon_{p}-1)/\epsilon_{p}.
\label{Q_P1}
\end{equation}

Now we apply this to our model: the total {\it polarisation} current element $Q_P$ is the sum on the polarisation current elements on
all the capacitors {\it in dielectric} $C_{d\,i}$. The contribution from each capacitor is $j\omega C_{d\,i}$ times the voltage
on this capacitor $Vf_i$, times the projection factor $g_i$, times $d_{d\,i}(\epsilon_r-1)/\epsilon_r$,
hence we obtain
\begin{align}
Q_P=&\sum_{i=1}^N j\omega f_i V g_i C_{d\,i}d_{d\,i}(\epsilon_r-1)/\epsilon_r  = \notag \\
    &j\omega \epsilon_0(\epsilon_r-1) A_0 V\sum_{i=1}^N g_i f_i
\label{Q_P}
\end{align}
Now comparing (\ref{Q_P1}) with (\ref{Q_P}), yields
\begin{equation}
\epsilon_{eq}\frac{\epsilon_{p}-1}{\epsilon_{p}} = (\epsilon_r-1)\frac{1}{N} \sum_{i=1}^N g_i f_i.
\label{compare_Q_P1}
\end{equation}
We divide it by $\epsilon_{eq}-1$ from Eq.~(\ref{epsilon_eq_1}), obtaining
\begin{equation}
\frac{(\epsilon_{p}-1)/\epsilon_{p}}{(\epsilon_{eq}-1)/\epsilon_{eq}} = \frac{\sum_{i=1}^N g_i f_i}{\sum_{i=1}^N f_i}
\label{compare_Q_P3}
\end{equation}
or in a more suggestive form:
\begin{equation}
\frac{(\epsilon_{p}-1)/\epsilon_{p}}{(\epsilon_{eq}-1)/\epsilon_{eq}} = \langle g \rangle.
\label{compare_Q_P2}
\end{equation}
where $\langle g \rangle$ is the projection factor averaged by the fraction of voltage in the
dielectric. Given that $0\le g_i\le 1$, also 
\begin{equation}
0\le\langle g \rangle\le 1,
\label{g_range}
\end{equation}
and hence
\begin{equation}
1\le\epsilon_{p}\le\epsilon_{eq}=n_{eq}^2,
\label{epsilon_av_range}
\end{equation}
however it seems that $\langle g \rangle$ cannot be 0 for a physical system, hence the lower limit
should be bigger than 0, so that practically $\epsilon_{p}>1$ always. Hence we consider the case of $\langle g \rangle=0$,
or $\epsilon_{p}=1$ only in the context of ``ignoring the transverse polarisation''. For the microstrip example (see Figure~\ref{quiver_ms}), $\langle g \rangle\simeq 1$, so that
$\epsilon_{p}\simeq\epsilon_{eq}$, but for the circular shaped conductors cross section (see Figure~\ref{quiver_cylinders}), 
$\langle g \rangle = 0.68$.

Using this model, we can also show that the solution of Eq.~(\ref{eps_av}) yields (\ref{compare_Q_P2}).
The integral in Eq.~(\ref{eps_av}), carried over the dielectric region, yields
on capacitor $i$, $V^+g_if_iA_0$, and this is summed on all capacitors:
\begin{equation}
\iint\limits_{\substack{\text{dielectric} \\ \text{region}}} dx' dy' E_x=A_0V^+\sum_{i=1}^N g_if_i.
\label{sum_Ex}
\end{equation}
Using $V^+=Z_0I^+$, Eq.~(\ref{Z0_1}), and $c\eta_0=1/\epsilon_0$, one obtains
\begin{equation}
\frac{\epsilon_{p}-1}{\epsilon_{p}}=\frac{\epsilon_0(\epsilon_r-1)A_0}{C d} \sum_{i=1}^N g_if_i,
\label{eps_av1}
\end{equation}
and using $C$ from Eq.~(\ref{total_C_1}), reproduces exactly Eq.~(\ref{compare_Q_P1}), leading to the result
(\ref{compare_Q_P2}).

Returning to the discussion at the beginning of this appendix (from which we derived Eq.~(\ref{alternative})),
we understand from Eq.~(\ref{compare_Q_P2}) that the physical meaning of $d_p$ is expressed by the relation
\begin{equation}
\frac{d_p}{d}=\langle g \rangle,
\label{alternative1}
\end{equation}
so that in the representation we used in this work of keeping the separation value $d$, the projection factor lies in the
definition of $\epsilon_{p}$. In the alternative representation of replacing $\epsilon_{p}$ by $\epsilon_{eq}$ and use for the effective
separation the value $d_p$, the projection lies in the separation $d_p$.



%

\end{document}